\documentclass[11pt]{article}
\usepackage{xcolor}
\usepackage{ulem}
\usepackage{dcolumn}
\usepackage{bm}
\usepackage{amsfonts}
\usepackage{amsbsy}
\usepackage{graphicx}
\usepackage{psfrag}
\usepackage{verbatim}
\usepackage{color}
\usepackage{amsmath}
\usepackage{epsfig,bm,epsf,float}
\usepackage{mathrsfs}
\usepackage{latexsym}
\usepackage{mathrsfs}
\usepackage{soul}
\usepackage{cancel}
\usepackage{newtxtext}
\usepackage{newtxmath}
\usepackage{natbib}
\usepackage{hyperref}
\usepackage{authblk}
\usepackage{blindtext}
\usepackage[a4paper, total={5.3in, 9.8in}]{geometry}

\usepackage[stdsubgroups]{nomencl} % Nomenclature
\makenomenclature

\usepackage{ifthen}
\renewcommand{\nomgroup}[1]{%
\ifthenelse{\equal{#1}{G}}{\item[\textbf{Greek Symbols}]}{%
\ifthenelse{\equal{#1}{S}}{\item[\textbf{Subscripts}]}{}}}

\graphicspath{{./fig/}}
\newcommand{\Reyn}{\text{\textit{Re}}}

\newcommand{\ie}{i.e.\ }
\newcommand{\ue}{\mathrm{e}}
\newcommand{\ui}{\mathrm{i}}

\newcommand{\ms}{\kern.10em\relax}
\graphicspath{{./fig/}}

\begin{document}
\title{Oscillating flow around a circular cylindrical post confined between two parallel plates}
\author[1,2]{Antonio J. Bárcenas-Luque\thanks{ajbarcenas@ugr.es}}
\author[3,4]{F. Moral-Pulido}
%\email{fmoral@ujaen.es} 
\author[3,4]{C. Guti\'errez-Montes}
%\email{cgmontes@ujaen.es}
\author[5]{W. Coenen}
%\email{wcoenen@ing.uc3m.es}
\author[1,2]{C. Mart\'inez-Bazán}
%\email{cmbazan@ugr.es}
\affil[1]{Department of Mechanics of Structures and Hydraulic Engineering, University of Granada, 18001 Granada, Spain.}
\affil[2]{Andalusian Institute for Earth System Research, University of Granada, Avda. del Mediterr\'aneo s/n, 18006, Granada, Spain.}
\affil[3]{Department of Mechanical and Mining Engineering, University of Ja\'en, 23071 Ja\'en, Spain.}
\affil[4]{Andalusian Institute for Earth System Research, University of Ja\'en, Campus de las Lagunillas, 23071, Ja\'en, Spain.}
\affil[5]{Department of Thermal and Fluid Engineering, University Carlos III of Madrid, 28911 Legan\'es, Spain.}

\date{}
\maketitle

\begin{abstract}
This work is motivated by the interest in determining the effect of the micro-anatomy of the spinal subarachnoid space on the cerebrospinal fluid flow and on the associated transport of solutes. To that aim, we focus on a canonical model problem in which a circular post of radius $a$, confined between two parallel plates separated by a distance $2h$ of order $a$, is subjected to an oscillatory flow. In particular, we are concerned with the steady, viscous, time-averaged flow that persists in the vicinity on the cylinder, when 
the stroke length of the oscillating flow, $U_{\infty}/\omega$ is smaller or comparable to the radius of the post, $a$, where $U_{\infty}$ and $\omega$ are the mean velocity amplitude and the corresponding angular frequency, respectively.
First, we analyze the asymptotic limit of small values of the stroke length, for a harmonic waveform, varying the aspect ratio of the post $\lambda$ and the Womersley number $M=\left(a^2 \omega/\nu\right)^{1/2}$. First-order steady-streaming corrections of the harmonic flow at leading order have been computed, together with associated Stokes-drift component, the sum of both yielding the mean Lagrangian velocity field. The confinement effect is seen to induce the three-dimensionality of flow. Moreover, similarly to the two-dimensional case, in the mid plane the time-averaged steady flow exhibits a recirculating vortex attached to the wall of the post that decreases as $M$ increases for low values of $M$. However, for values of $M$ larger than a critical one, $M_{cr}(\lambda)$, that depends on $\lambda$, a second, outer vortex is also formed. The dependence $M_{cr}(\lambda)$ has been quantified in the range $\lambda$ from 0.5 to $\infty$, where $M_{cr}$ increases as $\lambda$ decreases. The analysis has been corroborated experimentally for a fixed aspect ratio, $\lambda=2$, and varying the Womersley number, the stroke length and the wave form of the oscillating flow. The experiments show similar flow patterns when the stoke length is increased from $\varepsilon =U_{\infty}/\omega a= 0.2$ to 0.5, with more intense vortices at larger values of $\varepsilon$. Consideration of an anharmonic oscillating flow shows the fort-and-aft symmetry of the steady flow is broken, with the formation of two vortices of different size when $M< M_{cr}$ and only one outer vortex in the systolic direction when $M>M_{cr}$. Finally, the analysis is experimentally extended to consider an array of equally spaced posts, separated a semi-distance $d=2a$ aligned with the flow axis of oscillation.
\end{abstract}

\section{Introduction}
\begin{figure}
\centering
\includegraphics[width=0.5\linewidth, angle=0]{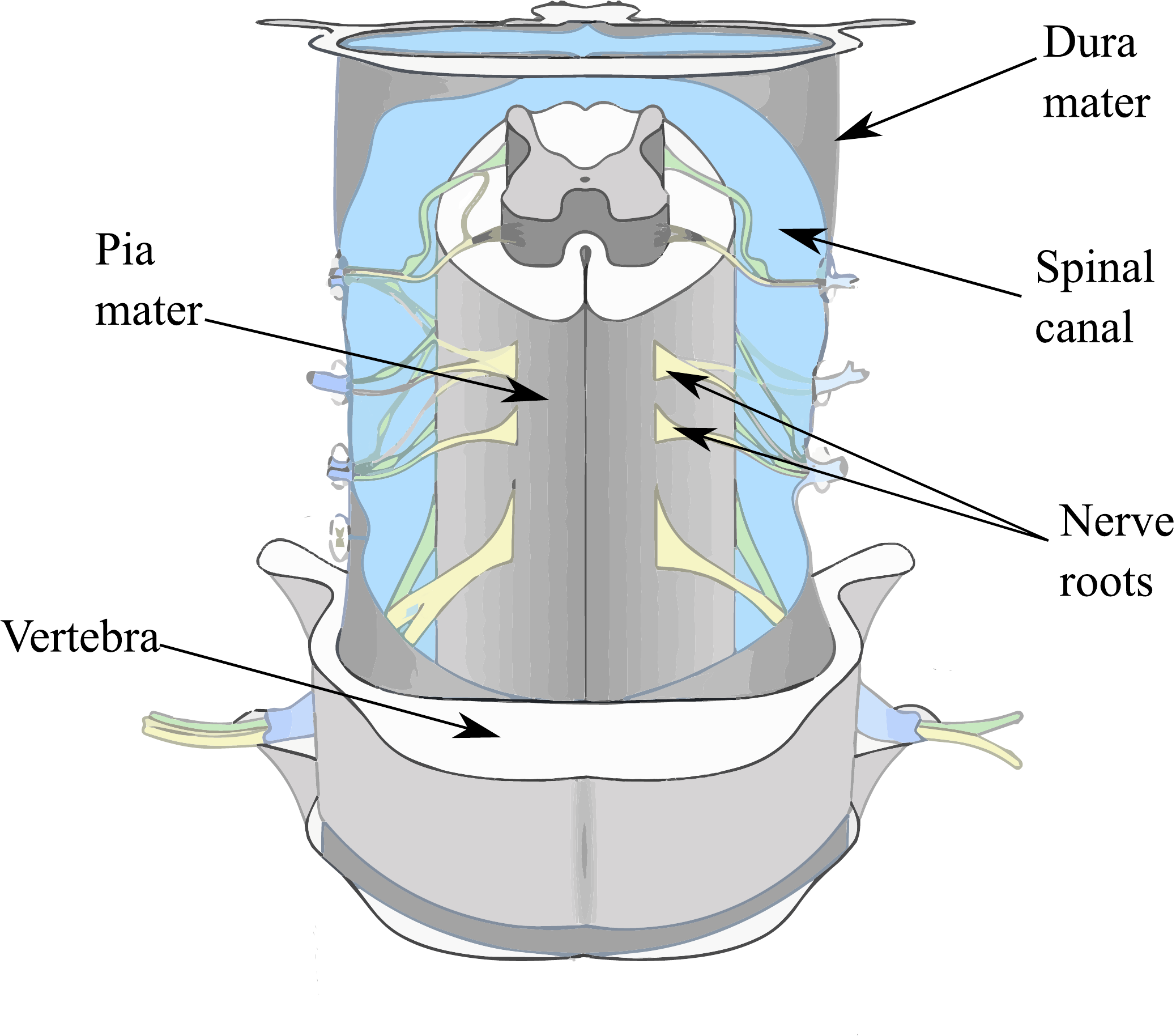}
\caption{Representation of the spinal canal with indication of the nerve roots across it.}
\label{fig:microfeatures}
\end{figure}

\begin{figure}
 \centering
\includegraphics[width=0.75\linewidth]{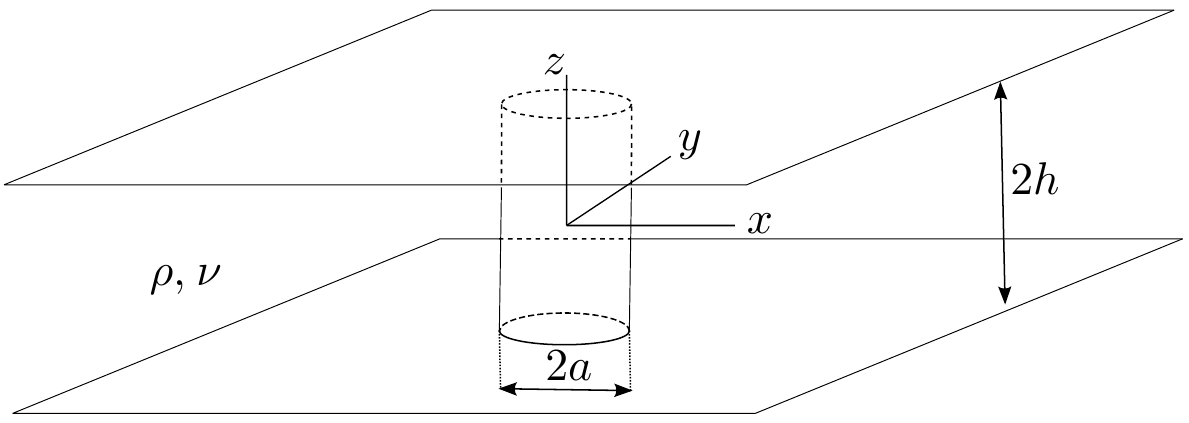}
    \caption{Schematic configuration of the flow.}
    %\caption{Schematic configuration of the flow around a cylindrical post, confined between two parallel plates.}
    \label{fig:model}
\end{figure}

The spinal subarachnoid space (SSAS) forms an annular slender canal filled of cerebrospinal fluid (CSF) bounded externally by the dura membrane and internally by the pia membrane surrounding the spinal cord. Along it, CSF features an oscillatory velocity, mainly induced by the combined contribution of the cardiac and the respiratory cycles~\cite{Linninger2016,Kelley2023}. The anatomy of the SSAS is known to present a complex geometry, including the presence of microanatomical elements~\cite{Dauleac2019}, namely, trabeculae, denticulate ligaments and nerve roots, across the width of the canal, that is much smaller than its perimeter and which in turn is much smaller than its length. In particular, nerve roots are present in most spinal segments and can be found on both sides of the canal arranged in bundles that emerge from the spinal cord and exit the vertebral column through the intervertebral discs (see Fig.~\ref{fig:microfeatures}). These elements have been reported to play an important role in the transport of substances along the spinal canal~\cite{stockman2006effect, stockman2007effect,tangen2015,pahlavian2014impact,Reina2020,Ayansiji2023determination}, such as drugs injected intrathecally. 
The present paper, focused on a simplified geometry, is inspired by the study of the effect of such elements on the flow of CSF and the transport of solutes in the SSAS. Thus, taking into account the anatomy of the canal, the description of the effect of nerve roots in the SSAS can be approached, in a first approximation, as that of a flat canal in the presence of obstacles facing the oscillatory CSF flow, see Fig.~\ref{fig:model}. This configuration is used herein to elucidate the influence of these anatomical elements on the CSF flow, and more specifically on the mean, steady Lagrangian motion that is generated, and which is mainly responsible for the transport of solutes~\cite{paper2}. The mean Lagrangian velocity experienced by a fluid particle in the SSAS is the sum of the steady-streaming velocity, determined by time-averaging the Eulerian velocity field~\cite{riley2001steady, paper1}, and the Stokes drift \cite{Stokes1847}, a purely kinematic effect associated with the spatial nonuniformity of the pulsatile flow.\\ 

The phenomenon of steady or acoustic streaming, documented by \cite{Rayleigh1883} in the last quarter of the 19th century, has been extensively studied since then and it is considered a classical fluid mechanics problem. It is known that an oscillating fluid stream, characterized by a velocity $U_\infty \cos(\omega t')$, interacting with a stationary solid body, leads to a time-averaged steady-streaming motion \citep{riley2001steady}. The resulting solution depends on the velocity amplitude, $U_\infty$, the object size, $a$, the oscillating angular frequency, $\omega$, and the fluid's kinematic viscosity, $\nu$. These parameters can be combined to give two dimensionless, controlling parameters, i.e the dimensionless stroke length $\varepsilon=U_\infty/\omega a$, and the Womersley number $M=\left(a^2 \omega/\nu\right)^{1/2}$, related to the Reynolds number by $\Reyn=U_\infty a/\nu=\varepsilon M^2$. For $\varepsilon\ll 1$, the problem permits a theoretical description, where the velocity components are expressed as an asymptotic expansion involving powers of $\varepsilon$. The leading-order terms, satisfying convection-free linear equations, manifest as harmonic functions with zero time-averaged values, while the first-order corrections introduce a non-zero steady-streaming component \citep{riley2001steady}.\\

In the case of two-dimensional oscillating flow around a circular cylinder of radius, $a$, much smaller than its length, \cite{holtsmark1954boundary} derived an analytical description of the Eulerian velocity for $\varepsilon \ll 1$, reformulated in primitive variables in Appendix~\ref{appendix}. This analysis provided expressions for both the dominant harmonic velocity and the subsequent first-order velocity corrections, later refined by ~\cite{chong2013inertial}. In the distinguished regime of $M \sim 1$, the magnitude of the resulting steady-streaming velocity is of the order of that of the Stokes drift~\cite{raney1954acoustical}, what  requires to take into account both phenomena to fully describe the mean, steady Lagrangian flow. In this case, the symmetry of the problem yields identical recirculatory patterns in all four quadrants. More precisely, when $M$ is smaller than a critical value, $M_{cr}$, each quadrant displays a single vortex directed towards the cylinder along the oscillation axis. However, for larger values of $M$ an additional outer vortex is formed, a feature also confirmed by accompanying experiments~\citep{holtsmark1954boundary}. As $M$ increases, the outer vortex strengthens while the inner one diminishes, being ultimately confined to a thin near-surface Stokes layer when $M \gg 1$. Similar flow characteristics have been observed when the flow is confined between two coaxial cylinders, with the inner one oscillating and the outer one fixed~\citep{skavlem1955coaxial}.\\

In the early seventies, ~\cite{tatsuno1973} investigated experimentally the secondary flow induced by an oscillating circular cylinder confined radially for values of $M \sim 1$ and finite dimensionless stroke lengths $\varepsilon$, finding a behaviour similar to that of ~\cite{skavlem1955coaxial} but different values of the critical parameters. While the oscillating flow for $\varepsilon \ll 1$ remains periodic and symmetric about the oscillation axis, the solution encountered when $\varepsilon$ takes values that are not sufficiently small is known to be more complicated. The periodic viscous flow becomes unstable to axially periodic vortices above a critical value of $\varepsilon$ that depends on $M$ leading to an asymmetrical flow featuring vortex shedding. This symmetry breaking is apparent in the experiments of Tatsuno and Bearman \citep{tatsuno_bearman_1990}, who identified eight different regimes of flow and reported a three-dimensional instability along the axis of the cylinder for most of the regimes with turbulent motion arising as the Reynolds number $\Reyn = \varepsilon M^2$ exceeded a critical value. Furthermore, Justesen \citep{justesen_1991} carried out a numerical study of planar oscillating flow around a circular cylinder in the range of high Reynolds number and finite $\varepsilon$ where, depending on the parameters values, several flow regimes that included vortex shedding were documented.\\

Regarding three-dimensional streaming flows in confined geometries, flow visualization has shown that, in the distinguished limit $M \sim 1$ with solid boundaries, depending on the boundary layer thickness, the confined case remains very similar to the two-dimensional one towards the central plane
\citep{lutz2005microscopic,lutz2006}. In the same regime, measurements of the three-dimensional velocity field show apparent three-dimensional effects which cannot be ignored, except in the midplane where symmetry enforces no axial motion, i.e. null velocity in the cylinder's axis direction \citep{tien2013}. Also for $M \sim 1$ and streaming Reynolds numbers, $\Reyn_s=\varepsilon ^2 M^2 \ll 1$, when the boundary surface is a bubble excited acoustically confined between two parallel walls, experiments reveal that there are large axial displacements near the bubble \citep{marin2015three}, which was also analytically reported later \citep{rallabandi2015three}. Similarly, three-dimensional streaming flows involving multi-curvature bodies have been also studied~\citep{chan_2022}, as well as configurations where a series of cylinders are placed transverse to~\cite{coenen2016} or aligned with~\cite{Alaminos-Quesada2023} the oscillating flow, among others.\\ 

The objective of the present work is to delve into the complexities of an oscillating, viscous and incompressible flow around a circular cylindrical post confined between two parallel flat plates, as depicted in Fig.~\ref{fig:model}. As mentioned above, the work is motivated by the analysis of the effect on the CSF flow in the spinal canal of existing microanatomical features that bridge irregularly the SSAS, see Fig.~\ref{fig:microfeatures}.  In this case, taking into account that the radii of the nerve roots range between $a\sim 0.25- 2~\textup{mm}$~\cite{Liu2015}, and the width of the canal is about $h\sim 1- 2.5~\textup{mm}$~\cite{coenen2019subject, sass20173d}, their aspect ratio, defined by $\lambda = h/a$, is about $\lambda\sim 1-10$ and the Womersley number $M\sim 1$. To the best of our knowledge, situations where the obstacle is confined between parallel plates, with $\varepsilon\ll 1$, $M \sim 1$ and the aspect ratio $\lambda$ is varied have not been explored. \\

The paper is organized as follows. In Section~\ref{formulation}, an standard asymptotic analysis of steady-streaming flows for small stroke lengths, $\varepsilon \ll 1$, has been carried out together with the computation of the velocity correction or Stokes-drift. The numerical results are presented in Section \ref{numerics}. In particular, the influence of the aspect ratio of the post on the streaming flow and the evaluation of the Lagrangian mean velocity field are reported in this Section. Section \ref{experiments} is devoted to describe the complementary experiments performed, the results obtained being reported in Section~\ref{experimental_results}. Finally, concluding remarks, limitations and future work are presented in Section \ref{conclusions}.\\

\section{Formulation}\label{formulation}
A fluid of density $\rho$ and kinematic viscosity $\nu$ fills the gap between two infinite parallel plates separated by a distance $2h$ (see Fig.~\ref{fig:model}). Application of a uniform harmonic pressure gradient of angular frequency $\omega$ in a given direction $x$ parallel to the plates results in a unidirectional motion with harmonic velocity $\textbf{v}'$given by
\begin{equation}\label{womersley}
    \frac{\textbf{v}'}{u_c}= \textup{Re} \left\{\ui\ue^{\ui\omega t'}\frac{\textup{cosh}\left[\left(\frac{1+\ui}{2}\right)h/\sqrt{\nu/\omega}\right]-\textup{cosh}\left[\left(\frac{1+\ui}{2}\right)z'/\sqrt{\nu/\omega}\right]}{\textup{cosh}\left[\left(\frac{1+\ui}{2}\right)h/\sqrt{\nu/\omega}\right]-1}\right\}e_x
\end{equation}
where $t'$ and $z'$ are the dimensional time and distance to the central plane, respectively, $e_x$ is the unit vector parallel to the pressure gradient, and $u_c$ is the peak velocity amplitude, occurring at $z'=0$. The exact solution (\ref{womersley}) is altered by the presence of obstacles. We consider in particular
the case of a circular cylinder perpendicular to the plates whose radius $a$ is comparable to
the inter-plate semidistance $h$, so that $\lambda=h/a\sim1$. The description assumes frequencies $\omega$ comparable to the inverse of the viscous time $(a^2/\nu)^{-1}$ and stroke lengths $u_c/\omega$ small compared with the cylinder radius $a$, as measured by the parameters
\begin{equation}
    M=\left(\frac{a^2\omega}{\nu}\right)^{1/2}\sim 1 \,\,\,\ \textup{and} \,\,\,\ \varepsilon=\frac{u_c/\omega}{a}\ll1.
\end{equation}
Using $\omega$, $a$, $u_c$ and $\rho u_c\omega a$ as scales for the time $t$, cartesian coordinates $(x,y,z)$, velocity $\textbf{v}=(u,v,w)$, and spatial pressure difference $p$ reduces the problem to that of integrating
\begin{equation}\label{continuity}
    \nabla\cdot\textbf{v}=0,
\end{equation}
\begin{equation}\label{momentum}
    \frac{\partial\textbf{v}}{\partial t}+\varepsilon\textbf{v}\cdot\nabla\textbf{v}=-\nabla p+ \frac{1}{M^2}\nabla^2\textbf{v},
\end{equation}
for $x^2+y^2>1$ and $-\lambda\leq z\leq\lambda$ subject to the nonslip boundary conditions
\begin{equation}\label{nonslip}
    \textbf{v}=0 \left\lbrace \begin{array}{ll}
     \textup{at } x^2+y^2=1 \,\,\ \textup{for } -\lambda\leq z\leq\lambda \\
     \textup{at } z=\pm \lambda \,\,\ \textup{for } x^2+y^2>1
    \end{array}
    \right.
\end{equation}

and the far-field condition
\begin{equation}\label{far-field}
    \textbf{v}=\left(\textup{Re}[\ui\ue^{\ui t}U_0(z)],0,0\right) \,\ \textup{as} \,\ x^2+y^2\rightarrow\infty \,\ \textup{for} \,\ -\lambda\leq z\leq +\lambda,
\end{equation}
where
\begin{equation}
    U_0(z)=\frac{\textup{cosh}\left(\frac{1+\ui}{2}M\lambda\right)-\textup{cosh}\left(\frac{1+\ui}{2}Mz\right)}{\textup{cosh}\left(\frac{1+\ui}{2}M\lambda\right)-1},
\end{equation}
as follows from Eq.~(\ref{womersley}). In the limit $\varepsilon\ll 1$, the dependent variables can be expressed as expansions in powers of $\varepsilon$ as,
\begin{equation}
    \textbf{v}=\textbf{v}_0+\varepsilon\textbf{v}_1+\cdots \,\,\ \textup{and} \,\,\ p=p_0+\varepsilon p_1+\cdots
\end{equation}
As can be anticipated from the two dimensional configuration, at leading order in the limit $\varepsilon\ll 1$ the resulting motion is determined by a linear problem involving the balance of the local acceleration with the pressure and viscous forces. Convective acceleration will be seen to introduce a small relative correction of order $\varepsilon$, including a steady-streaming component, which is to be determined here.\\

\subsection{Leading-order flow}\label{subsec:leading-order}

The leading-order harmonic solution can be expressed in the form $\textbf{v}_0 = \textup{Re}[\ui\ue^{\ui t}\textbf{V}]$ and $p_0 = \textup{Re}[\ue^{\ui t}P]$, where $\textbf{V}(x,y,z) = (U,V,W)$ and $P(x,y,z)$, by integration of
\begin{equation}\label{leading.order.continuity}
    \nabla\cdot\textbf{V}=0,
\end{equation}
\begin{equation}\label{leading.order.momentum}
    -\textbf{V}=-\nabla P+\frac{\ui}{M^2}\nabla^2\textbf{V},
\end{equation}
subject to the nonslip boundary conditions
\begin{equation}\label{Lo.bc1}
    \textbf{V}=0 \left\lbrace \begin{array}{lll}
     \textup{at } x^2+y^2=1 \,\,\ \textup{for } -\lambda\leq z\leq\lambda \\
     \textup{at } z=\pm \lambda \,\,\ \textup{for } x^2+y^2>1
    \end{array}
    \right.
\end{equation}

and the far-field condition
\begin{equation}\label{Lo.bc2}
    \textbf{V}=[U_0(z),0,0] \,\,\ \textup{as} \,\,\ x^2+y^2\rightarrow\infty \,\,\ \textup{for} \,\,\ -\lambda\leq z\leq +\lambda.
\end{equation}

\subsection{Steady streaming}\label{subsec:Steady-streaming}
Collecting terms of order $\varepsilon$, the following order becomes
\begin{equation} \label{v1_cont}
    \nabla\cdot\textbf{v}_1=0,
\end{equation}
\begin{equation}\label{v1_momentum}
    \frac{\partial\textbf{v}_1}{\partial t}+\textbf{v}_0\cdot\nabla\textbf{v}_0=-\nabla p_1+ \frac{1}{M^2}\nabla^2\textbf{v}_1,
\end{equation}
with boundary conditions
\begin{equation}\label{vo1_bc}
    \textbf{v}_1=0 \left\lbrace \begin{array}{lll}
     \textup{at } x^2+y^2=1 \,\,\ \textup{for } -\lambda\leq z\leq\lambda \\
     \textup{at } z=\pm \lambda \,\,\ \textup{for } x^2+y^2>1\\
     \textup{as } x^2+y^2\to\infty\,\,\ \textup{for } -\lambda\leq z\leq\lambda
    \end{array}
    \right.
\end{equation}
The first-order corrections $\textbf{v}_1$ and $p_1$ are $2\pi$-periodic functions of time. As a result of the nonlinear interactions associated with the convective terms, their time-averaged values $\textbf{v}_{SS}=\langle \textbf{v}_1 \rangle$ and $p_{SS}= \langle p_1 \rangle$ are nonzero, with $\langle \cdot \rangle =\int_0^{2\pi}\cdot dt $ denoting the time-average operator. Taking the time average of (\ref{v1_cont})-(\ref{vo1_bc}) and using the identity $\langle\textbf{v}_0\cdot\nabla\textbf{v}_0\rangle=\frac{1}{2}\textup{Re}[\textbf{V}\cdot\nabla\textbf{V}^{*}]$, where $*$ denotes complex conjugates, yields the steady-streaming problem
\begin{equation} \label{ss_cont}
    \nabla\cdot\textbf{v}_{SS}=0,
\end{equation}
\begin{equation}\label{ss_momentum}
    \frac{1}{2}\textup{Re}[\textbf{V}\cdot\nabla\textbf{V}^{*}]=-\nabla p_{SS}+ \frac{1}{M^2}\nabla^2\textbf{v}_{SS},
\end{equation}
with boundary conditions
\begin{equation}\label{ss_bc}
    \textbf{v}_{SS}=0 \left\lbrace \begin{array}{lll}
     \textup{at } x^2+y^2=1 \,\,\ \textup{for } -\lambda\leq z\leq\lambda \\
     \textup{at } z=\pm \lambda \,\,\ \textup{for } x^2+y^2>1\\
     \textup{as } x^2+y^2\to\infty\,\,\ \textup{for } -\lambda\leq z\leq\lambda
    \end{array}
    \right.
\end{equation}

The steady-streaming velocity $\textbf{v}_{SS}=\langle \textbf{v}_1 \rangle$ provides the leading-order description for the
mean Eulerian velocity $\langle \textbf{v} \rangle = \varepsilon \textbf{v}_{SS}$ in the asymptotic limit $\varepsilon \ll 1$.

\subsection{Lagrangian mean motion}
When addressing the oscillating flow around a cylinder, the Lagrangian mean motion of the fluid particles comes partly from the contribution of the Eulerian mean motion ($\textbf{v}_{SS} = \langle\textbf{v}_1\rangle$) and partly from that of the so-called Stokes drift, arising in non-uniform oscillating flows. Consequently, streamlines visualized in experiments employing dyed fluid do not coincide in general with those determined from the steady-streaming velocity. The velocity of the Lagrangian mean
motion is,
\begin{equation}
    \textbf{v}_L=\textbf{v}_{SS}+\frac{1}{2}\textup{Im}(\textbf{V}\cdot\nabla\textbf{V}^{*})
\end{equation}
where $\frac{1}{2}\textup{Im}(\textbf{V}\cdot\nabla\textbf{V}^{*})$ corresponds to the Stokes drift contribution. %It is of interest to highlight that the real part of the complex function $\frac{1}{2}\textbf{V}\cdot\nabla\textbf{V}^{*}$ determines the steady streaming, as revealed by (\ref{ss_cont}), (\ref{ss_momentum}), whereas its imaginary part is the Stokes-drift velocity. Note that, as mentioned before, for large values of $M$ viscous forces are confined to a thin Stokes layer, outside of which the flow is potential and the function $\textbf{V}$ is real, so that the associated Stokes drift can be expected to vanish for $M\gg 1$.
Of particular interest is the role played by  the complex function $\frac{1}{2}\textbf{V}\cdot\nabla\textbf{V}^{*}$ in determining the Lagrangian mean motion. In the steady streaming formulation, as discussed in (\ref{ss_cont}) and (\ref{ss_momentum}), the forcing term is the real part of the complex function, while its imaginary part gives the Stokes-drift velocity. Notably, in the scenario of large $M$, viscous forces predominates within a thin Stokes layer. Beyond this layer, the flow is potential, characterized by the function $\textbf{V}$ being real. Consequently, one can anticipate a that the contribution of the Stokes drift is negligible in comparison with the steady-streaming one for large values of $M$.
\begin{figure}
 \centering
\includegraphics[width=0.8\linewidth]{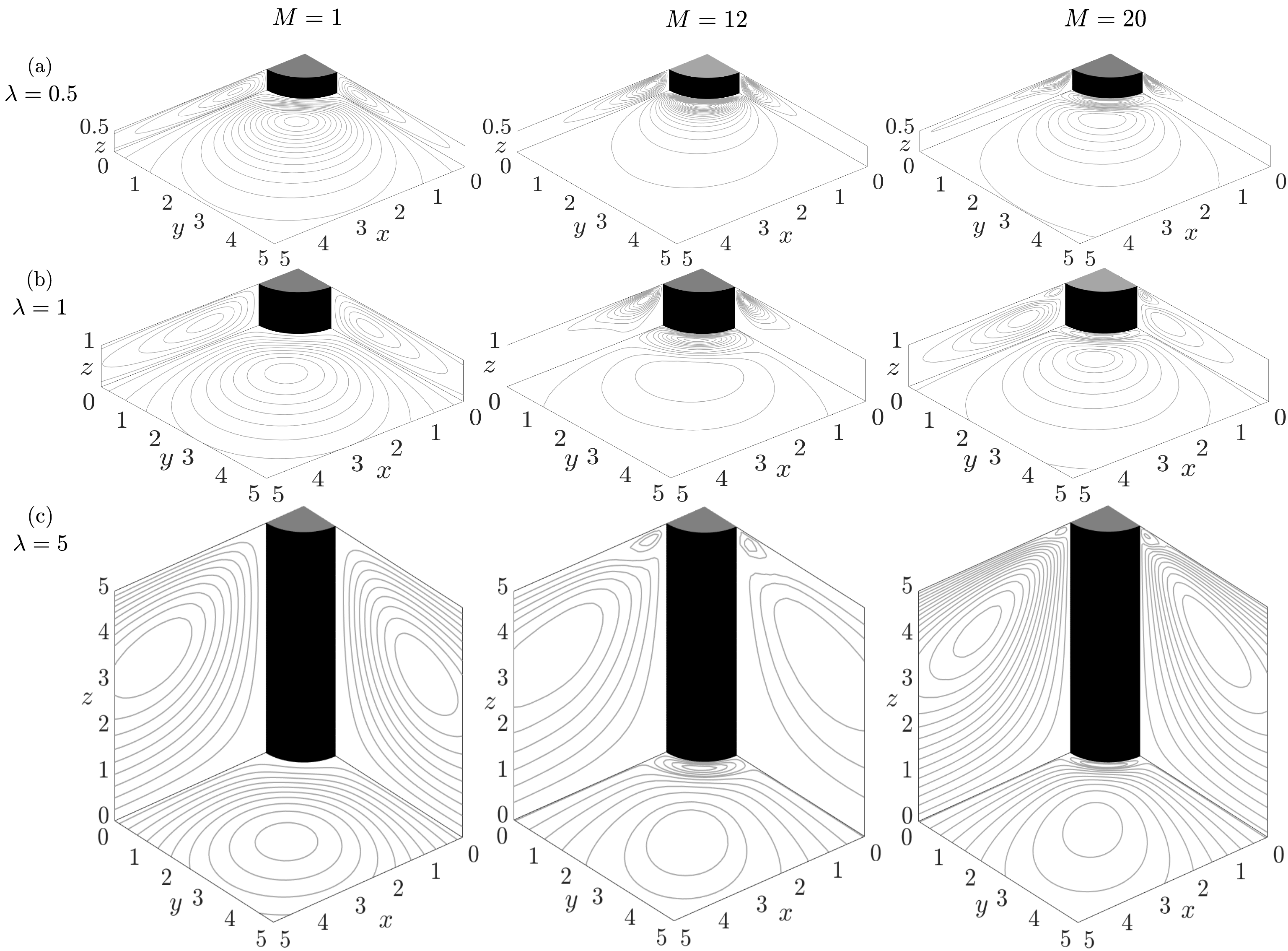}
    \caption{Streamlines in the symmetry ($z = 0$, $y = 0$) and anti-simmetry ($x = 0$) planes corresponding to the steady-streaming velocity, $\textbf{v}_{SS}$, for $\lambda = 0.5$ (a), $\lambda = 1$ (b) and $\lambda = 5$ (c) for three different values of $M$.}
    \label{fig:3d}
\end{figure}
\section{Numerical results}\label{numerics}
For $\lambda \sim 1$, no analytic solution is available, and the problem formulated in Section~\ref{formulation} must be solved numerically. To that aim, the Eqs.~(\ref{leading.order.continuity}), (\ref{leading.order.momentum}) subject to (\ref{Lo.bc1}), (\ref{Lo.bc2}) and Eqs.~(\ref{ss_cont}), (\ref{ss_momentum}) subject to (\ref{ss_bc}) were written in weak form and implemented in the finite element solver 
COMSOL Multiphysics\textsuperscript{\tiny\textregistered} v5.6 using the Weak Form PDE toolbox. In the regime herein investigated, i.e. $M\sim 1-10$ and $\varepsilon \ll 1$, the flow can be anticipated to be symmetric respect to the planes $y=0$ and $z=0$, and, thus, the computational domain for $y\geq1$ and $z\ge1$ has been considered imposing symmetry conditions in the aforementioned planes. The domain was discretized using a structured mesh with hexahedral elements. Second-order Lagrange elements were used for the pressure and the velocity. Mesh elements were compressed towards the solid and symmetry boundaryes to assure enough spatial resolution. Both the domain extension and the mesh resolution were varied to assure the domain-size and grid-resolution independence of the results. In the final configuration, the typical element size ratio with respect to the radius, $a$, was ranged from $10^{-3}$ at the surfaces to $10^{-1}$ near the far-field boundary, assuring the accuracy of the solution with reasonable computational times. The domain extension in the $x$- and $y$-directions had a length of twenty-five times the radius of the cylinder, \ie $25a$.

\subsection{Steady streaming}
\begin{figure}
 \centering
\includegraphics[width=1\linewidth]{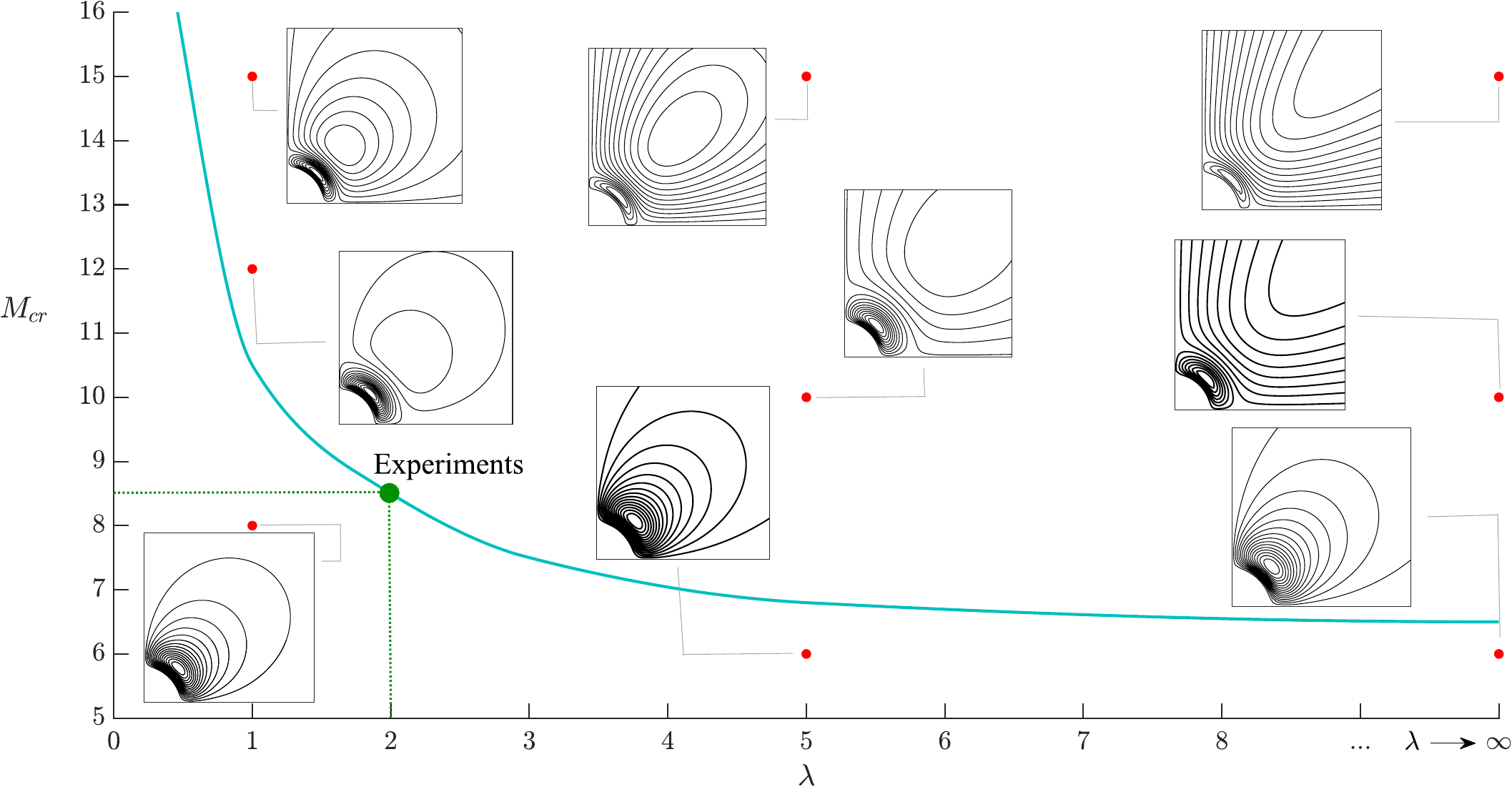}
    \caption{Critical value of the Womersley number, $M_{cr}$, as a function of $\lambda$. Insets show streamlines in the central plane, $z = 0$, for different values of $M$ and $\lambda$. The green point indicates the value of $M_{cr}$ corresponding to the experimental configuration reported in Section~\ref{experiments}.}
    \label{fig:Mcr}
\end{figure}

There are different methods to represent and visualize a three-dimensional flow field. Some of these methods can be found in the literature (see, for instance \cite{jeong_hussain_1995} and \cite{katsanoulis_2023}). In the cases of jet or wake-like flows, the $Q$-criterion or the $\lambda_2$-criterion are widely accepted for the representation of the flow through vortical structures. When the exact analytical solution can be found, streamsurfaces are often preferred to characterize the flow. In the particular case at hand, the absence of the normal component of the velocity field in the symmetry, $z = 0$ and $y = 0$, and antisymmetry planes, $x = 0$, and the fact that the symmetry of the flow is conserved when $\varepsilon \ll 1$, allows the visualization of the flow topology by means of representation of the steady streaming streamfunction contours in these planes. Only one component of the vorticity field is not zero in each coordinate plane, so the corresponding streamfuction comes from $\nabla^2\psi_{SS}^i = \Omega_{SS}^i$ for $i = 1 (z = 0),2 (y = 0),3 (x = 0)$ with homogeneous Dirichlet boundary conditions in the far-field, in the cylinder and in the walls.\\

Representative results are shown in Fig.~\ref{fig:3d} for $M=$ 1, 12 and 20, and for $\lambda=$ 0.5, 1 and 5. Because of the flow conditions and the geometrical configuration, the flow is symmetric respect to the plane $x=0$. Therefore, only half of the domain ($x \ge 0$) is shown. The streaming structure arising for finite values of the dimensionless aspect ratio $\lambda$ in the central plane $z = 0$ is qualitatively similar to that of a single cylinder ($\lambda \gg 1$). In this plane, the mean secondary Eulerian flow displays a vortex in each quadrant for the lower value of the Womersley number $M = 1$ (plotted in the left column), being the core of the vortex located along the $\pi/4$ ray and closer to the cylinder as $\lambda$ decreases. This vortex is also known to progressively approach the post wall as $M$ increases, as can be seen in Fig.~\ref{fig:3d} for $M = 1$ (left column), $M = 12$ (central column) and $M = 20$ (right column). These particular values of $M$ have been selected to show the transition in the topology of the flow for the three values of  $\lambda$, where a second vortex with opposite circulation begins to form outside when $M$ exceeds a critical value, $M_{cr}$. Thus, the critical value of the Womersley number, $M_{cr}$, represents a threshold for which a change in the topology of the flow is observed. \\

As seen in Fig.~\ref{fig:3d}, the critical value of the Womersley number depends on the aspect ratio $\lambda$. For $\lambda = 0.5$ (Fig.~\ref{fig:3d}a), it can be seen that the transition occurs between $M = 12$ and $M = 20$, while for $\lambda = 1$ and $\lambda = 5$ the transition happens for $M < 12$ (Figs.~\ref{fig:3d}b and c). More details will be given below. In the planes $y = 0$ and $x = 0$, a similar behaviour can be observed. As it happens in the plane $z = 0$, the vortex in each quadrant of these planes progressively approaches to the post as $M$ increases. The core of these vortices is located at $z = \lambda/2$ for low values of $M$ and it is situated at lager values of $z$, closer to the confining walls, when $M$ increases. That indicates that the flow tends to be two-dimensional near the central plane ($z=0$) for large values of $M$ and for large values of $\lambda$. When the Womersley number, $M$, is increased sufficiently, a Stokes layer appears close to the solid boundaries and in the limit of $\varepsilon \ll 1$, $M \gg 1$,  $\Reyn_{s} \gtrsim O(1)$, the flow outside this boundary layer becomes potential.\\  

\begin{figure}
 \centering
\includegraphics[width=0.85 \linewidth]{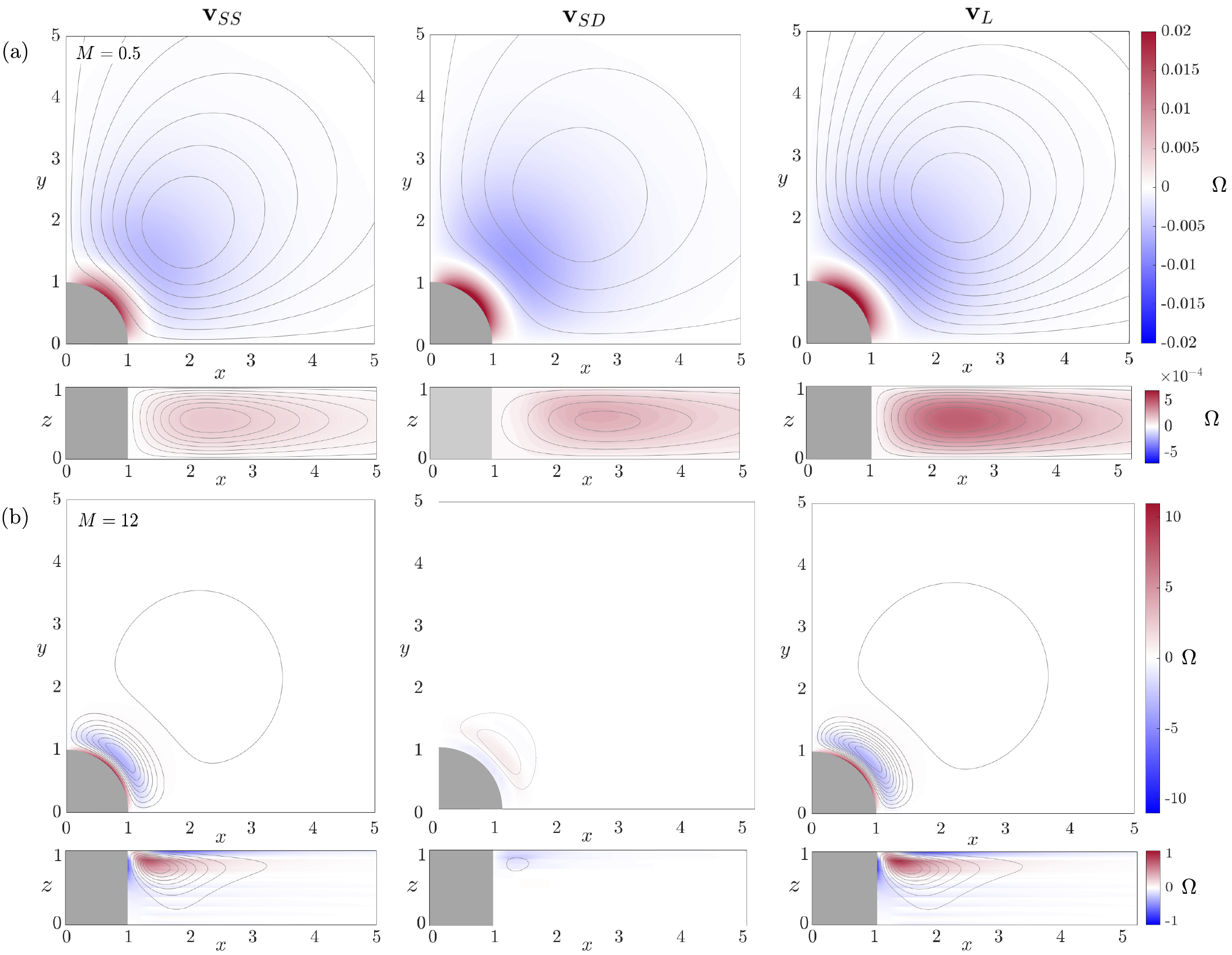}
    \caption{Streamlines and colour contours of vorticity $\Omega$ corresponding to the steady-streaming velocity, $\textbf{v}_{SS}$,
Stokes-drift velocity, $\textbf{v}_{SD}$, and steady mean Lagrangian velocity, $\textbf{v}_{L}=  \textbf{v}_{SS} +  \textbf{v}_{SD}$ for  $\lambda= 1$ and $ M = 0.5 $ (a) and $ M = 12 $ (b). For each value of $M$, streamlines are represented using a constant spacing $ \delta\psi= 0.01$ ($M = 0.5$) and
$ \delta\psi = 0.03$ ($M = 12$), with the corresponding vorticity levels indicated in the colour bar on the right.}
    \label{fig:SS_SD_LAG}
\end{figure}

 As previously indicated, the value of $M_{cr}$ for which a change in flow topology occurs depends on $\lambda$. For the case of an infinite cylinder, $\lambda \gg 1$, the critical value of the Womersley number can be determined from the exact solution~\citep{holtsmark1954boundary} ($M_{cr} \approx  6.08$) as the value of $M$ for which the streamfunction $\psi_{SS}$ vanishes in the far field. Our numerical simulations indicate that $M_{cr}$ varies with $\lambda$, increasing as $\lambda$ decreases. This dependency is shown in Fig.~\ref{fig:Mcr}, together with accompanying contours of the steady streaming streaming functions in the central plane $z = 0$ for different pairs of $M$ and $\lambda$, shown in the insets. The value of $M_{cr}$ is found to vary significantly for low values of $\lambda$. These higher values of $M_{cr}$ for lower values of $\lambda$ are attributable to the effect of confinement, which also produces a drastic reduction in the magnitude of the streaming motion. In contrast, for sufficiently large values of $\lambda$, the value of $M_{cr}$ tends to that of the infinite cylinder (unconfined case). In this regard, as it can be seen for the cases of $\lambda = 1, 5, \infty$ with $M = 15$ in the three upper insets, the core of the outer vortex moves away from the post as the aspect ratio increases. Identical behaviour is observed in the rest of the cases. Figure~\ref{fig:Mcr} also shows by a green point the value of $M_{cr}$ corresponding to the experimental configuration reported in Section~\ref{experiments}.\\ 
 
\subsection{Evaluation of the Lagrangian mean velocity}
Figure~\ref{fig:SS_SD_LAG} shows streamlines and color contours of vorticity $\Omega$ corresponding to three different velocity fields for $\lambda = 1$ and two Womersley numbers, i.e. $M = 0.5$ and $M = 12$. The first column shows the steady-streaming velocity, $\textbf{v}_{SS}$, the second one the Stokes-drift or velocity correction, $\textbf{v}_{SD}$, and the last one the steady mean Lagrangian velocity, given by the sum of the two of them, $\textbf{v}_{L} = \textbf{v}_{SS} + \textbf{v}_{SD}$. Figure \ref{fig:SS_SD_LAG}(a) displays streamlines and vorticity contours of the velocity fields in the symmetry planes $z = 0$ and $y = 0$ for $M = 0.5$, while in Fig.~\ref{fig:SS_SD_LAG}(b) the same results are presented for $M = 12$.\\
\begin{figure*}
    \centering
    \includegraphics[width=0.95 \textwidth]{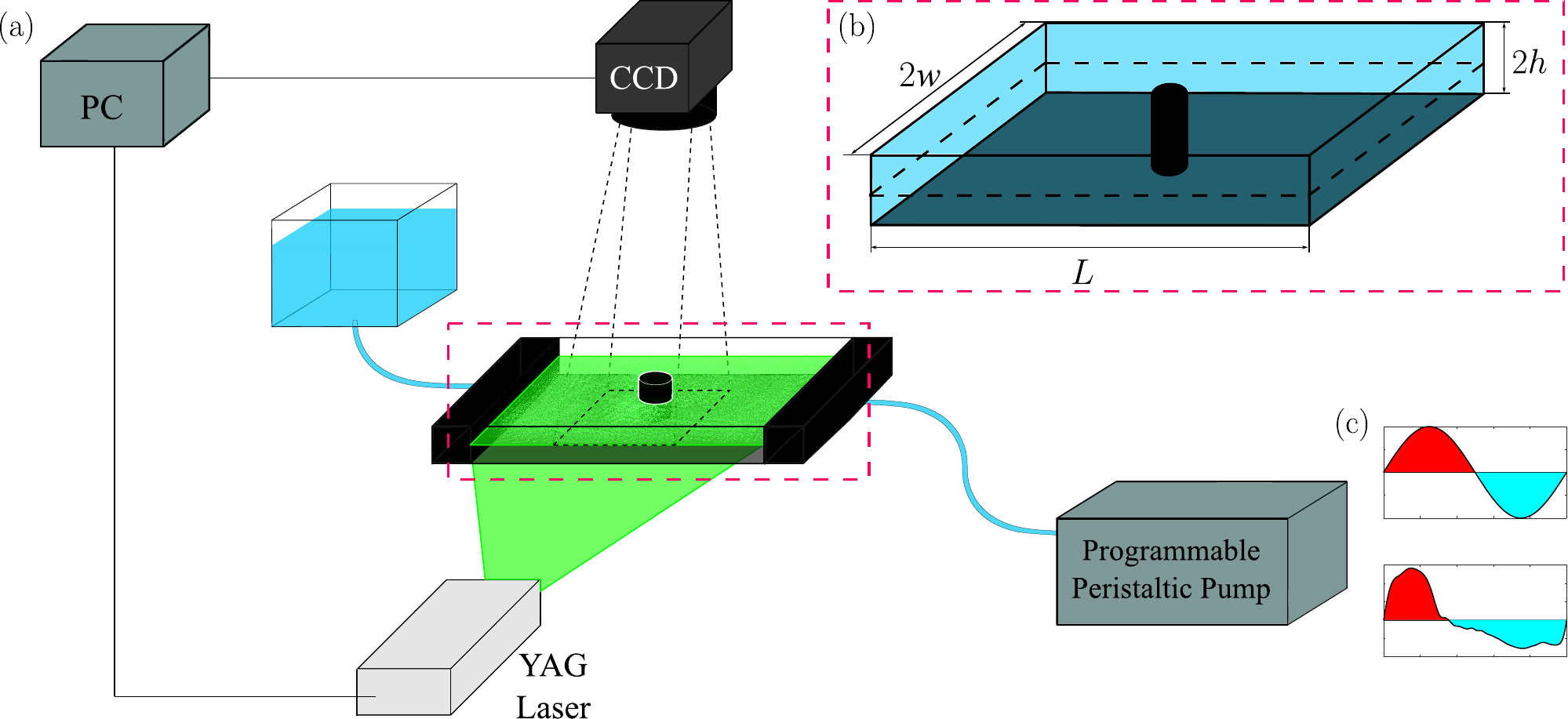}
    \caption{(a) Schematic view of the experimental facility, consisting of the in vitro model, the pumping system (programmable peristaltic pump and opened-air container) and the image aquisition setup (YAG laser and CCD camera, both synchronized with a PC). (b) In vitro model with the corresponding dimensions. (c) Flow rate waveforms used in the experiments.}
    \label{fig:exp_model}
\end{figure*}

Regarding the streaming structure in the plane $z = 0$ for $M = 0.5$, the flow displays one vortex in each quadrant, with the clockwise circulation (negative vorticity) exhibited by the vortex in the first quadrant corresponding to fluid approaching the cylinder along the oscillation axis $y = 0$. In the plane $y = 0$, the flow, that is much weaker than in the middle symmetry plane $z = 0$, displays also one vortex in the semi infinite space of $x > 1, y = 0$ with identical behaviour in the region of $x < -1, y = 0$. The Stokes-drift results for $M = 0.5$ display a primary clockwise-rotating vortex occupying most of the quadrant, along with a much weaker counter-rotating vortex of negligibly small circulation near the oscillation axis $y = 0$. For this value of $M$, this primary vortex is stronger than the corresponding steady-streaming vortex. The mean Lagrangian velocity field is largely determined by its Stokes-drift component, as reflected in the shape of the corresponding Lagrangian vortex in the right panel of Fig.~\ref{fig:SS_SD_LAG}(a).\\

As can be seen for $M = 12$ in Fig.~\ref{fig:SS_SD_LAG}(b), in the first quadrant of the center plane $z = 0$ the flow displays a second vortex with opposite circulation since $M$ exceeds the corresponding critical value $M_{cr}$ for $\lambda = 1$ ($M_{cr} \approx 10.2$), which in the case of Stokes-drift prevails over the inner vortex, similar to what happens in an array of cylinders \cite{Alaminos-Quesada2023}. The hegemony of the steady-streaming  over the Stokes-drift in the mean Lagrangian motion becomes apparent in this range of the Womersley number. As inferred from Fig.~\ref{fig:SS_SD_LAG}, the Stokes drift prevails for sufficiently small values of the Womersley number, whereas in the opposite limit the Stokes-drift motion fades away, as observed by comparing the steady-streaming and the mean Lagrangian motion contours, that are almost identical. Same dominance of the steady-streaming component happens in the plane $y = 0$, where no second vortex appears for this value of the Womersley number. \\

\section{Experiments}\label{experiments}
The complementary experiments carried out allowed us to analyse the long-time-scaled flow patterns of the motion generated for different flow conditions, varying the frequency, the stroke volume of the oscillating flow as well as the waveform. Two types of experiments were performed: the first one focused on the flow field around a single post and the second one on the flow around an array of 5 posts separated a distance $\ell =d/a$, where $a$ is the radius of the posts and $d$ the semi-distance between the axes of contiguous posts, being both types of experiments conducted in an adaptable experimental facility using distilled water of density $\rho = 998.2~\text{kg/m$^3$}$ and kinematic viscosity $\nu = 10^{-6}~\text{m$^2$/s}$ as working fluid. Figure~\ref{fig:exp_model}(a) shows a sketch of the complete experimental facility, with the test section illuminated with a laser sheet, which includes the programmable pump used to establish the oscillating flow and the acquisition system. As seen in Fig.~\ref{fig:exp_model}(b), the model consisted of an acrylic channel of total volume $V = 20 \times 10 \times 2~\text{cm$^3$}$, where $L = 20~\text{cm}$, $2w = 10~\text{cm}$ and $2h = 2~\text{cm}$ are the length, width and height of the channel, respectively. The bottom of the channel was covered with black adhesive vinyl, which acted as a background for the experiments and at the same time prevented laser reflections. A vertical cylindrical post of radius $a = 5 ~\text{mm}$ and length $2h = 2~\text{cm}$ was perpendicularly placed at the center of the canal, being the corresponding aspect ratio $\lambda = h/a = 2$. Such radius was sufficiently small to avoid any influence from the lateral confining walls on the fluid motion near the post, being the distance between the axis of the post and the walls equal to $w$ = 10$a$. This model configuration was also extended to consider a linear array of 5 posts, which were arranged aligned with the fluid motion. These elements were separated a distance $2d = 20~\rm{mm}$, being the dimensionless inter-post semi-distance $\ell = d/a = 2$. The \textit{in vitro} model was connected to a programmable peristaltic pump on one side, which generated the oscillatory flow with a prescribed waveform, and an open-air container on the other side, to allow the displacement of the stroke volume (see Fig.~\ref{fig:exp_model}a). The pump allowed to vary the relevant parameters of the flow: the stroke volume, $\Delta V$, the oscillatory frequency, $f=\omega/(2 \pi)$, and the waveform (from a harmonic sinusoidal wave to a MRI-based anharmonic one as displayed in Fig.~\ref{fig:exp_model}c). The fluid entered and exited the canal through two nozzles placed at the inlet and outlet sections, which also included flow conditioning elements to generate a uniform flow.\\

The experiments aimed at characterizing the time-averaged flow, induced by the oscillatory stream in the presence of the obstacles. To that end, in order to visualize the flow patterns and to measure the velocity and vorticity fields, particle image techniques were applied, seeding the flow with hollow glass spheres of diameter $\phi_p \approx 10 \, \upmu \rm{m}$. The experiments consisted of series of long-duration sequences of black-and-white images acquired in phase, at the same frequency as the oscillating flow. The sequences contained between 100 and 200 images depending on the experimental set, which were acquired with a CCD camera of resolution $2048\times 2048 ~\rm{px}$. The camera was synchronized with a pulsed YAG laser, Fig.~\ref{fig:exp_model}a), used to generate a horizontal planar green sheet ($\lambda_{\textup{laser}} = 541 ~\rm{nm}$) of width $\delta h < 1 ~\rm{mm}$ at $z = 0$.\\

The resulting images were analysed in two different ways to determine the fluid particle trajectories as well as the Lagrangian steady velocity and vorticity fields. To obtain the long-time-scaled Lagrangian trajectories, the images were first enhanced subtracting their background which was calculated using an average filter. Afterwards, the steady fluid flow patterns could be extracted integrating all the snapshots acquired in each experimental set, to yield a cumulative image. It should be emphasized that all the images were taken at the same phase at a sampling frequency, $f_s$, equal to that of the oscillating flow, $f$. In addition, the velocity and vorticity fields were also determined using Particle Image Velocimetry (PIV) analysis, applying the MATLAB toolbox PIVlab GUI~\cite{Stamhuis2014} to couples of images acquired at an interval of time, $\Delta t= 1/f$. To perform the analysis, the images were first pre-processed, masking the cylinder to eliminate solid regions and applying a high-pass filter of window size larger than 150 px to avoid non-uniform background illumination, in combination with an automatic contrast adjustment. The PIV routine was applied to the pre-processed images, using an initial window of $160 \times 160 ~\text{px}$ with a window reduction up to a minimal size of $32 \times 32 ~\text{px}$ in 4 steps, applying a 50\% window overlapping.\\

A series of experiments were conducted first to evaluate the  effects of the Womersley number, $M$, and the dimensionless stroke volume,  $\varepsilon = \left(U_{\infty}/\omega\right)/a$ where $U_{\infty}$ is the amplitude of the oscillating flow, on the flow characteristics using a post with a fixed value of the aspect ratio $\lambda = 2$, complementing the numerical study reported in Section~\ref{numerics}. Thus, a value $\varepsilon = 0.2$ was initially established, assuming that it was small enough to approximate the distinguished limit $\varepsilon \ll 1$, and the Womersley number was varied from $M = 5$ to 11 (experimental sets 1-7 and 13-19 in Table~\ref{tab:PA_exp}). Some of these experiments will be compared with the numerical results reported in Section~\ref{numerics}. Then, the stroke volume was increased to $\varepsilon=0.5$ to evaluate the effect of $\varepsilon$ (experimental sets 8-12 and 20-23 in Table~\ref{tab:PA_exp}). The experiments were performed considering two different waveforms in order to investigate the effect of the shape of the flow rate on the induced flow patterns. The first waveform consisted of an harmonic function that provided a periodic flow rate given by $Q'(t') = Q_{\textup{max}}\cos{\left(\omega t'\right)}$, or similarly, a periodic velocity far from the post  $u'_{\infty}(t') = U_{\infty}\cos(\omega t') \,$ with $\omega = 2\pi f$, $u'_{\infty}(t')= Q'(t')/(4wh)$ and $U_{\infty}= Q_{\textup{max}}/(4wh)$ (experimental sets 1-12 in Table~\ref{tab:PA_exp}). For the second waveform, a realistic anharmonic flow rate, obtained by MRI measurements, was used. This waveform, emulating the CSF motion along the spinal canal, was obtained by a modal Fourier analysis, using 8 modes to produce a well fitted signal as described in Section~\ref{effect_MRI}  (experimental sets 13-23 in Table~\ref{tab:PA_exp}). Finally, additional experiments, not included in Table~\ref{tab:PA_exp},  were conducted with a linear array of elements, aligned with the flow as described in Section~\ref{effect_array}.
\begin{table}
    \begin{center}
\def~{\hphantom{0}}
    \begin{tabular}{cccccc}
       Experiment & Waveform & $f$ (Hz) & $M$ & $U_{\infty}$ (mm/s) & $\varepsilon$ \\[3pt]
        1 & Sine & 0.16 & 5 & 1.005 & 0.2 \\
        2 & Sine & 0.24 & 6 & 1.508 & 0.2 \\
        3 & Sine & 0.32 & 7 & 2.01 & 0.2 \\
        4 & Sine & 0.40 & 8 & 2.51 & 0.2 \\
        5 & Sine & 0.50 & 9 & 3.14 & 0.2 \\
        6 & Sine & 0.64 & 10 & 4.02 & 0.2 \\
        7 & Sine & 0.78 & 11 & 4.90 & 0.2 \\[3pt]
        8 & Sine & 0.16 & 5 & 2.51 & 0.5 \\
        9 & Sine & 0.24 & 6 & 3.77 & 0.5 \\
        10 & Sine & 0.32 & 7 & 5.03 & 0.5 \\
        11 & Sine & 0.40 & 8 & 6.28 & 0.5 \\[3pt]
        12 & Sine & 0.16 & 9 & 7.85 & 0.5 \\[3pt]
        13 & MRI & 0.16 & 5 & 1.005 & 0.2 \\
        14 & MRI & 0.25 & 6 & 1.508 & 0.2 \\
        15 & MRI & 0.32 & 7 & 2.01 & 0.2 \\
        16 & MRI & 0.40 & 8 & 2.51 & 0.2 \\
        17 & MRI & 0.50 & 9 & 3.14 & 0.2 \\
        18 & MRI & 0.625 & 10 & 4.02 & 0.2 \\
        19 & MRI & 0.80 & 11 & 4.90 & 0.2 \\[3pt]
        20 & MRI & 0.16 & 5 & 2.51 & 0.5 \\
        21 & MRI & 0.25 & 6 & 3.77 & 0.5 \\
        22 & MRI & 0.32 & 7 & 5.03 & 0.5 \\
        23 & MRI & 0.40 & 8 & 6.28 & 0.5 \\
   \end{tabular}
    \caption{Experimental conditions of the different sets of the experiments performed. Here, $M = \left(a^{2}\omega/\nu\right)^{1/2}$, $f=\omega/(2 \pi)$, $U_\infty$ is the peak velocity of the CSF motion for the harmonic waveform, and the harmonic equivalent peak velocity of the CSF motion for the anharmonic one, and $\varepsilon = \left(U_{\infty}/\omega\right)/a$. `Sine' indicates that the wave form is harmonic and `MRI' that is an anharmonic cardiac signal obtained from Magnetic Resonance Imaging.}
    \label{tab:PA_exp}
    \end{center}
\end{table}
\begin{figure*}
    \centering
    \includegraphics[width = \textwidth]{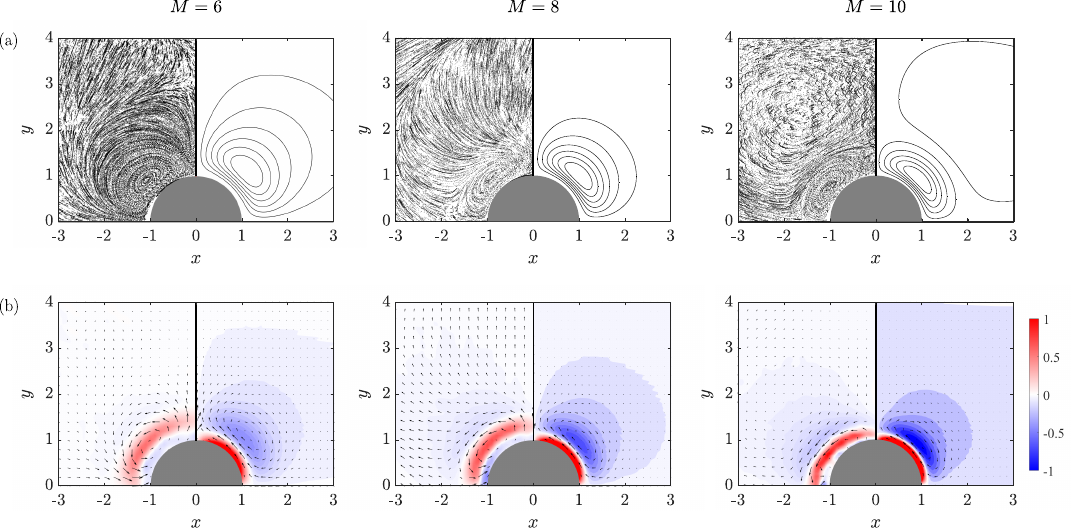}
    \caption{ (a) Steady pathlines and (b) velocity field and vorticity contours of the flow around a single cylindrical post for $\varepsilon = 0.2$ and different values of $M$. Left side of each panel corresponds to the experimental measurements and right side to the analytical predictions ($\varepsilon \ll 1$) }
    \label{fig:Figure3}
\end{figure*}
\section{Experimental Results}\label{experimental_results}
In this section, the results of the time-averaged Lagrangian flow induced by the interaction of an oscillating stream with a single post, and an array of posts, obtained from the \textit{in vitro} experiments are described. For the case of a single post, Section~\ref{effect_M} first discusses the effect of the Womersley number, $M$, for small values of the stroke length, $\varepsilon \simeq 0.2$, to be compared with numerical predictions of Section~\ref{numerics}, obtained for the limit $\varepsilon \ll 1$. These results are extended to order unity values of $\varepsilon$ in Section~\ref{effect_eps}, for which the asymptotic description in powers of $\varepsilon$, for $\varepsilon\ll 1$, is no longer valid. The analysis in the aforementioned sections is conducted assuming a harmonic sinusoidal wave. In Section~\ref{effect_MRI}, this study is extended to analyze the time-averaged flow induced by an anharmonic oscillatory wave, similar to that found in realistic CSF flows, as induced by intracranial pressure fluctuations generated by the cardiac cycle. Finally, the time-averaged lagrangian flow originated by the presence of an array of posts aligned with the flow oscillation axis is briefly described in Section~\ref{effect_array}.

\subsection{Effect of flow frequency, $M$}\label{effect_M}
As described in Section~\ref{formulation}, for small values of $\varepsilon\ll1$, the motion can be decomposed into the sum of a zero-averaged purely oscillatory velocity field and a first-order correction, whose non-zero average is denoted steady-streaming velocity. The Eulerian velocity field of the two-dimensional flow, $\lambda =\infty$, is described in Appendix~\ref{appendix}. Such description, although strictly valid for $\varepsilon \ll 1$, has been recently reported to provide results that remain reasonably accurate for finite values of $\varepsilon$, close to unity, for a linear array of circular cylinders~\cite{Alaminos-Quesada2023}. Furthermore, the description of the long-time scale motion of a fluid particle includes the additional contribution of the Stokes drift \cite{chong2013inertial} that, in combination with the steady streaming, yields the mean Lagrangian velocity field. The Stokes drift has been found to be comparable to the steady streaming for $M\leq1$, but negligible otherwise \cite{alaminos2021}. In the case at hand, since the values of $M$ experimentally considered vary between 5 to 11, it is expected that the time-averaged Eulerian and Lagrangian velocity fields almost coincide in the cases reported in the present section for $\lambda=2$.\\

In general terms, the results reveal that, as happens in the two-dimensional case, the Lagrangian mean motion in the horizontal symmetry plane $z=0$ shows identical recirculation patterns in the four quadrants. Consequently, for the single post case, results in only one quadrant will be reported in the following, see Fig.~\ref{fig:Figure3}. In this configuration, two different flow topologies can be identified. For values of the Womersley number below a critical one (denoted hereafter as sub-critical values, $M<M_{cr}$), a sole vortex forms in each quadrant (left panels in Fig.~\ref{fig:Figure3}, $M=$ 6), which, in the first quadrant, rotates in clockwise direction, i.e. the flow moves towards the post along the stream's oscillation axis, $y=0$, and away from the post along the antisymmetry axis, $x=0$. For values larger than the critical one (or super-critical values), $M>M_{cr}$, a secondary external counter-rotating vortex appears, that, differently from what happens in the two-dimensional flow, in this case is closed due to the presence of the confining walls (right panels in Fig.~\ref{fig:Figure3}, $M$= 10). The external vortex encloses the internal one close to the wall of the post at distances that decrease as $M$ increases. As indicated in Section~\ref{numerics}, the presence of the plates at the top and bottom boundaries has also an impact on the critical value, $M_{cr}$, at which such flow topology change takes place. In particular, $M_{cr}$ increases as $\lambda=h/a$ decreases, being $M_{cr}\simeq 6.08$ in the infinite cylinder case, $\lambda = \infty$, and $M_c\simeq8.5$ in the present configuration, $\lambda = 2$ (see Fig.~\ref{fig:Mcr}), which practically coincides with the value obtained analytically for $\varepsilon \ll 1$. In addition, the numerical analysis shows that the time-averaged flow becomes three-dimensional, and it is characterized by the formation of two additional recirculating regions in the planes $x=0$ and $y=0$, although this issue will not be experimentally explored, due to optical access limitations. To illustrate these observations, Fig.~\ref{fig:Figure3} shows the comparison between the time-averaged analytical ($\varepsilon\ll1$) and experimental ($\varepsilon\simeq0.2$) results. In particular, Fig.~\ref{fig:Figure3}(a) shows pathlines of the mean Lagrangian motion obtained experimentally, left or second quadrant, and analytically, right or first quadrant, for values of $M=$ 6, 8 and 10, in columns from left to right, respectively. The first two columns display flow topologies corresponding to $M<M_{cr}$, whereas a change in the flow topology becomes apparent for $M=10$, where a second external counter-rotating vortex appears, which confines the inner vortex to the high-vorticity Stokes layer close to the post. Figure~\ref{fig:Figure3}(b) shows the vectors associated to the velocity fields, together with colour contours of vorticity, normalized using the overall maximum analytical and experimental values of velocity and vorticity, respectively. In general terms, very similar and symmetric flow patterns are found for both the theoretical predictions and the measurements, with the analytical or experimental cores of the vortices located, approximately, along the $\pi/4$ or $3\pi/4$ rays, respectively, in all cases, the theory predicting slightly larger radial distances from the post to the vortex core. Finally, it can be also observed that increasing values of $M$ yield more intense circulations, as inferred from the vorticity contours, while the core of the vortex, for $M<M_{cr}$, and that of the internal vortex, for $M>M_{cr}$, become closer to the post, as it was previously anticipated.\\ 
\begin{figure}[t]
    \centering
    \includegraphics[width = 0.85\textwidth]{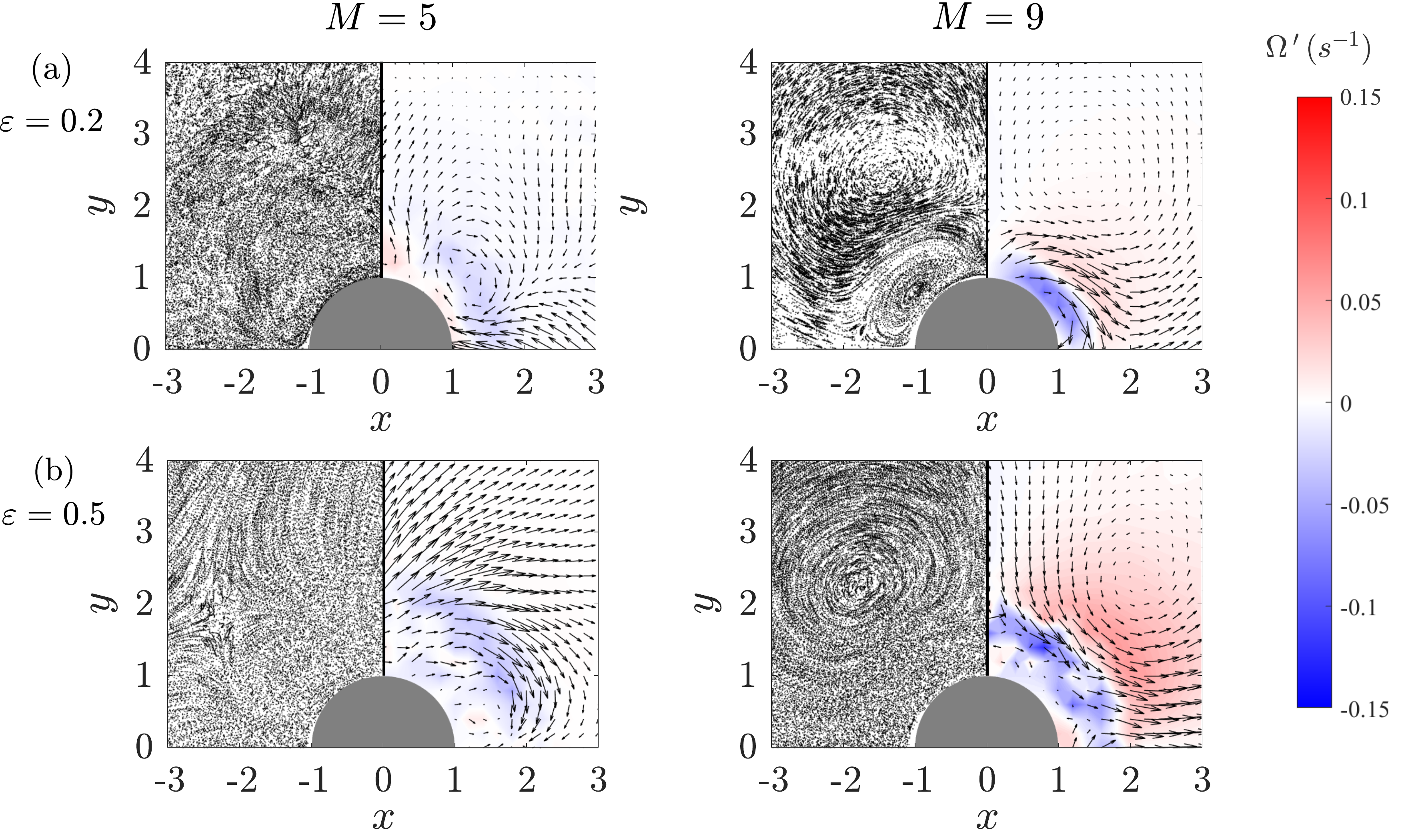}
    \caption{Steady flow patterns (left side of each panel) and corresponding dimensional velocity and vorticity fields (right side of each panel) of the flow around a cylindrical element for $M < M_{cr}$ (left column, $M = 5$) and $M > M_{cr}$ (right column, $M = 9$). Here, a) $\varepsilon= 0.2$ is considered to satisfy the theoretical approximation ($\varepsilon \ll 1$) and b) $\varepsilon= 0.5$ assumes to be of order-unity.}
    \label{fig:Figure4}
\end{figure}
\subsection{Effect of $\varepsilon$}\label{effect_eps}

The flow rate and stroke volume varies along the spinal canal, affecting the local value of $\varepsilon$, which also varies from order unity values at the cervical region to very small values at the lower regions. The asymptotic description in powers of $\varepsilon$ in the limit $\varepsilon \ll 1$ ceases to be valid for finite values of the stroke length, $\varepsilon\sim\mathcal{O}(1)$. To explore this effect, the experimental results for $\varepsilon\sim\mathcal{O}(1)$ are shown herein for two values of $M$, namely $M = 5 < M_{cr}$ (sub-critical regime) and $M = 9 > M_{cr}$ (super-critical regime), which are also compared with their counterparts for $\varepsilon=0.2$ (considered here to be a good representation of the distinguished limit $\varepsilon \ll 1)$. Therefore, Fig.~\ref{fig:Figure4} displays pathlines of time-averaged Lagrangian flow, left or second quadrant, and the corresponding velocity and vorticity fields, right or first quadrant, for $M=5$, first column, and for $M=9$, second column, for $\varepsilon=0.2$ in Fig.~\ref{fig:Figure4}(a), and for $\varepsilon=0.5$ in Fig.~\ref{fig:Figure4}(b). In this figure, the vorticity contours are shown with dimensions, $\Omega'$ to facilitate the comparison. In both $M$ cases, similar flow patterns are observed for $\varepsilon=0.2$ and $\varepsilon=0.5$. Thus, a unique vortex is generated with clockwise rotation in the first quadrant for $M=5$, whereas for $M=9$, a second external counter-rotating recirculating region appears, enclosing the internal vortex near the post. A close inspection of the results reveals that increasing the stroke length has an impact on the global flow topology and rotation intensity. In that regard, for $M=9$, the size of the vortices decrease as $\varepsilon$ increases, and their cores move closer to the post. The latter is accompanied with more intense recirculations, as observed from the larger amplitudes of vorticity in the case of $\varepsilon=0.5$ compared to those of $\varepsilon=0.2$ in Fig.~\ref{fig:Figure4}. These results agree with those recently reported by \cite{Alaminos-Quesada2023}.

\begin{figure}[t!]
    \centering
    \includegraphics[width = 0.5\linewidth]{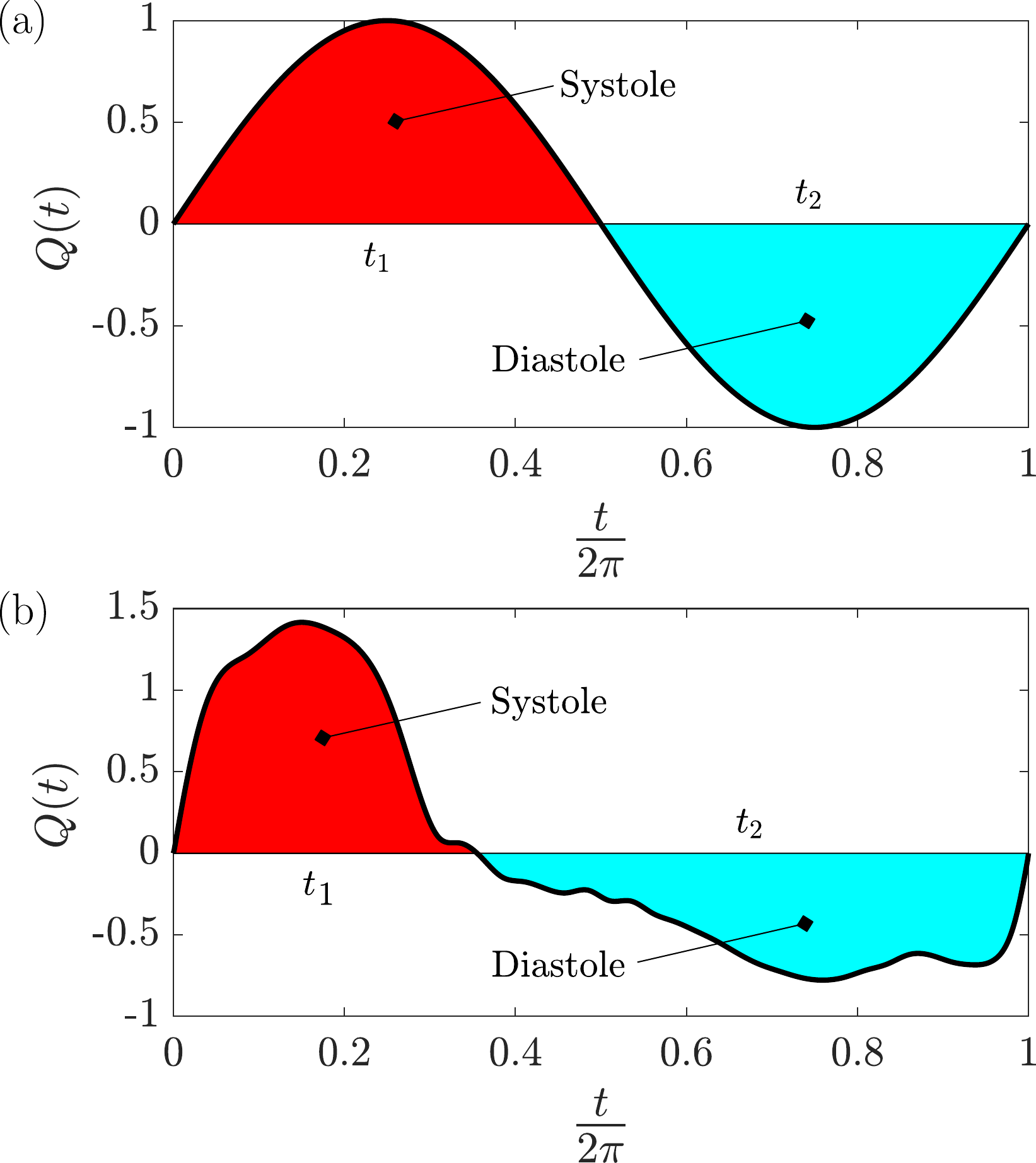}
    \caption{(a) Harmonic flow waveform ($t_1 = t_2 = \pi$). (b) Realistic flow waveform, obtained through MRI measurements and set as Fourier expansion ($t_1 \neq t_2$). Here, $t=2 \pi t'/T$ represents the dimensionless time, with $T=1/f=2\pi/\omega$ as the period of the waveform.}
    \label{fig:FigureQ}
\end{figure}
\begin{figure*}[t!]
    \centering
    \includegraphics[width = \textwidth]{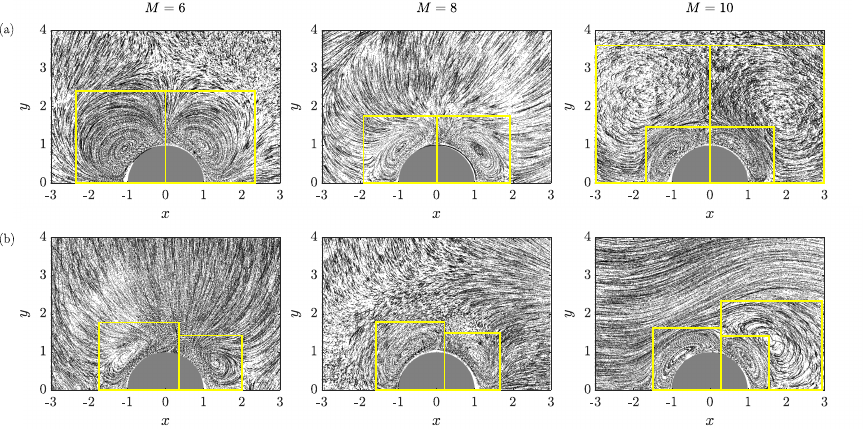}
    \caption{Steady flow patterns around a cylindrical post of aspect ratio $\lambda=2$ for $\varepsilon = 0.2$ and different values of $M$ using, a harmonic waveform (a) and a anharmonic one (b).}
    \label{fig:Figure5}
\end{figure*}

\subsection{Time-averaged Lagrangian motion in anharmonic oscillating flows: CSF flow}\label{effect_MRI}
So far the analysis has been focused on the flow around a post subject to a harmonic stream flow of velocity $u'_{\infty}(t')=U_{\infty}\cos{\omega t'} \,$. However, it is common to find oscillatory flows that present anharmonic temporal variations in nature, such as that of the blood in veins and arteries, or the CSF flow in the central nervous system (CNS)~\cite{Linninger2016, Kelley2023}. Therefore, in the present section, the study has been experimentally extended to analyze the flow induced by an anharmonic oscillatory flow around a post. Specifically, a wave of temporal dependence equal to that of a CSF flow measured in a human spinal canal~\cite{Sincomb2022} has been considered. This velocity waveform can be expressed as a Fourier series $u'_{\infty}(t')=\Sigma^{\infty}_{n=1}Re[A_n e^{i(\omega_n t' + \phi_n)}]$, where $A_n$, $\omega_n$ and $\phi_n$ are respectively the amplitude, angular frequency and phase associated to each mode \cite{Davidson1972, Alaminos-Quesada2023, alaminosquesadaAMMOD2023}. Figure~\ref{fig:FigureQ}(a) shows the flow rate associated to a harmonic waveform while Fig.~\ref{fig:FigureQ}(b) represents a period of the flow rate provided by a cardiac anharmonic waveform. In this figure, the positive and negative flow rates are given during systole, of dimensionless duration $t_1$, and diastole, of dimensionless duration $t_2$, respectively with $\int_0^{2 \pi} Q(t) dt=0$ or, similarly, $|\int_0^{t_1} Q(t) dt|=|\int_{t_1}^{2\pi} Q(t) dt|$, where $t=2 \pi t'/T$ represents the dimensionless time, and $T=1/f=2\pi/\omega$ is the period of the waveform. \\
\begin{figure*}[t!]
    \centering
    \includegraphics[width = 0.9\textwidth]{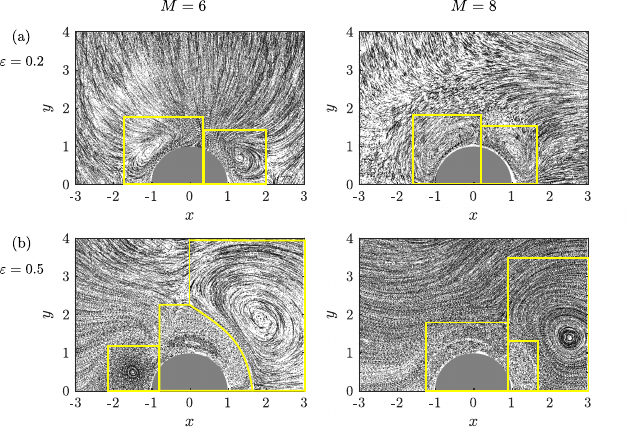}
    \caption{Steady fluid particles trajectories around a cylindrical post of aspect ratio $\lambda=2$ for $M= 6$ and 8, obtained with an anharmonic, subject-specific CSF waveform. (a) $\varepsilon$= 0.2 and (b) $\varepsilon$= 0.5.}
    \label{fig:Figure6}
\end{figure*}
\begin{figure*}[t!]
    \centering
    \includegraphics[width =1 \textwidth]{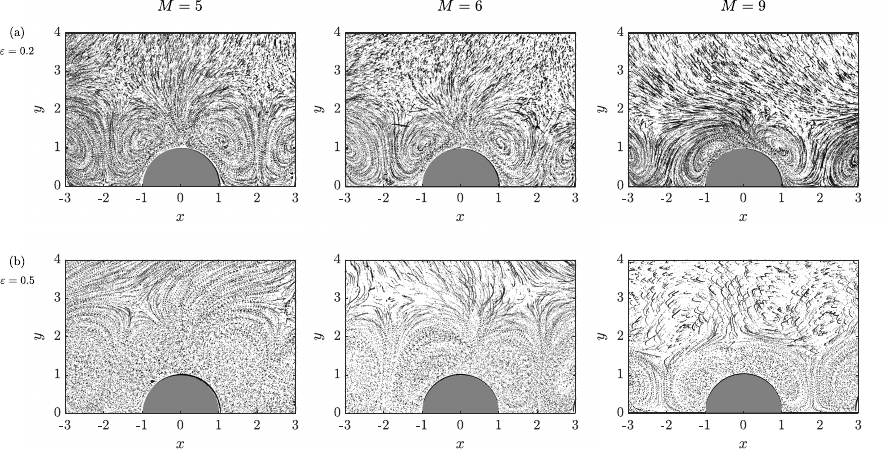}
    \caption{Steady flow patterns around a linear array of cylindrical posts separated a distance $\ell = 2$ at different values of $M$ (columns) induced by a harmonic waveform for (a) $\varepsilon$= 0.2 and (b) $\varepsilon$= 0.5.}
    \label{fig:Figure8}
\end{figure*}

Unlike the results reported in the previous sections for the harmonic waveform where the secondary steady flow patterns were doubly symmetric, the flow induced by an anharmonic wave results in a steady motion which is only symmetric with respect to the axis of the flow oscillation $y=0$. Consequently, in the following, the first two quadrants, $y\geq0$, of the plane $z=0$ will be shown to describe the flow. Figure~\ref{fig:Figure5}(a) shows pathlines of the time-averaged Lagrangian flow for the harmonic wave,  with their counterparts for the anharmonic wave shown in Fig.~\ref{fig:Figure5}(b), both for $\varepsilon=0.2$ and $M=$ 6, 8 and 10, respectively, where the first mode of the Fourier series, $\omega_1$,  was used to define $M$ in the anharmonic case and $U_\infty$ has been defined as that of the equivalent harmonic waveform signal with the same period and stroke volume. As previously shown, the harmonic wave induces steady flow patterns that are symmetric in each quadrant. However, this symmetry is broken in the anharmonic case, as a result of the inter-modes interaction \cite{Alaminos-Quesada2023, alaminosquesadaAMMOD2023}. For the lowest value of $M$ ($M=6$), the lack of fore-and-aft symmetry is already observable, where the vortices appear distorted, with their cores near the post, as in the harmonic case. Despite this fact, differences can be observed in the size of the vortices, with the vortex located in the positive midplane ($x>0$) being smaller than its complementary one in the negative one ($x<0$). The loss of symmetry is also apparent for larger values of $M$, e.g. $M=8$, where, as for $M=6$, two vortices form close to the post with their cores displaced downwards in the systolic direction (towards the positive $x$-coordinate). In addition, the two vortices seem to be surrounded by a region of fundamentally streamwise velocity, which becomes more noticeable for even larger values of $M$. It should be noted that, for $M=10$, the flow topology has already changed. However, differently from the flow pattern observed when a harmonic waveform is imposed, where a second outer vortex is formed in each quadrant, the steady flow is characterized by the formation of only an outer vortex in the first quadrant (downstream in the systolic direction). Moreover, the two vortices, already present for $M<M_{cr}$, undergo larger distortions and are also displaced downwards in the systolic direction. These three vortices are surrounded by an almost horizontal external stream, which is in agreement with the results obtained for the infinite cylinder \cite{tatsuno1981secondary} or more recently in a linear array of circular cylinders \cite{Alaminos-Quesada2023}.

The effect of considering finite stroke lengths is analyzed in Fig.~\ref{fig:Figure6}, where results for $\varepsilon=0.5$ (Fig.~\ref{fig:Figure6}b) are compared with those previously shown for $\varepsilon=0.2$ (Fig.~\ref{fig:Figure6}a), for $M=$ 6 and 8. As happened in the harmonic case, the size of the vortices formed is reduced when $\varepsilon$ is increased, these being also more distorted and displaced downwards in the systolic direction in this case. Interestingly, as it was previously anticipated, increasing $\varepsilon$ has also an impact on the flow topology change, reducing the value of $M_{cr}$. The latter can be observed for both values of $M$, for which only two vortices are formed close to the post when $\varepsilon=0.2$, one in each quadrant, whereas a third external vortex appears when $\varepsilon=0.5$ in the first quadrant already at $M=6$.

\subsection{Time-averaged motion around a linear array of posts}\label{effect_array}
\begin{figure*}
    \centering
    \includegraphics[width =1 \textwidth]{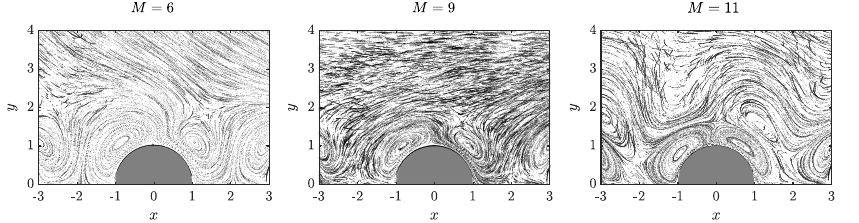}
    \caption{Steady flow patterns around a linear array of cylindrical posts separated a distance $\ell = 2$ for different values of $M$, induced by an anharmonic, patient-specific, waveform with $\varepsilon= 0.2$.}
    \label{fig:Figure9}
\end{figure*}
To further investigate the effect of the presence of different elements on the steady flow induced by an oscillatory stream flowing around them, the present section reports experimental results of the time-averaged Lagrangian flow around an array of five circular posts, aligned with the axis of oscillation, which are separated a semi-distance between the axes of contiguous posts $d$, with the dimensionless semi-distance $\ell=d/a=2$.\\

Figure~\ref{fig:Figure8} shows time-averaged pathlines in the horizontal symmetry plane $z=0$ generated by a harmonic flow for different values of $M$, namely $M=$ 5, 6 and 9, and two values of $\varepsilon$, i.e. $\varepsilon=$ 0.2 (Fig.~\ref{fig:Figure8}a) and 0.5 (Fig.~\ref{fig:Figure8}b), for $\ell=2$. Qualitatively, the steady flow patterns induced for $\ell=2$ are similar to those for $\ell = \infty$. Thus, considering the results for $\varepsilon=0.2$, for $M=5$ and 6, two symmetric, counter rotating vortices are generated, one in each quadrant, with the same direction of rotation as for $\ell=\infty$. The presence of contiguous posts renders a confining effect, limiting the shape of each vortex to the semi-distance length in the axial direction. Consequently, deformed vortices are generated, whose cores are displaced towards the antisymmetry axis ($x=0$), in agreement with \cite{Alaminos-Quesada2023}. This effect becomes more apparent for the lowest studied value of the Womersley number, $M=5$, since the vortices reduce their size and move closer to the posts when $M$ increases. The presence of neighbouring posts also has an impact on $M_{cr}$, delaying the flow transition featuring the apparition of external counter rotating vortices in each quadrant to larger values of $M$. The latter is noticeable for $M=9$, where only two internal vortices are formed, one in each quadrant, whereas its counterpart for $\ell=\infty$ displays two vortices in each quadrant ($M_{cr}\simeq8.5$ for $\ell=\infty$) as can be seen in Fig.~\ref{fig:Figure4}. Figure~\ref{fig:Figure8}(b) represents the results obtained for $\varepsilon=0.5$ for the same three values of $M$, also imposing a harmonic velocity. From an inspection of Fig.~\ref{fig:Figure8}, it cannot be concluded that the size of the vortices substantially change with $\varepsilon$ (at least in the range of values of $\varepsilon$ at which we have been able to perform the experiments). However, an increase of $\varepsilon$ decreases the value of $M_{cr}$, as it was commented in previous sections. This can be observed in the case of $M=9$, for which a pair of inner and outer counter rotating vortices are formed in each quadrant when $\varepsilon=0.5$, whereas only an inner vortex is seen when $\varepsilon=0.2$. The linear array configuration is further explored in Fig.~\ref{fig:Figure9}, where time-averaged pathlines induced by an anharmonic fluid stream are shown for three values of $M$, two of them in the sub-critical regime ($M=6~\text{and}~9$), and the other one in the super-critical regime ($M = 11$), all of them at $z=0$ and for $\varepsilon=0.2$. These kinds of waveforms have been found to cause the loss of flow symmetry in all quadrants, thus, this flow is only symmetric respect to the axis $y=0$. As a consequence, for $M<M_{cr}$, two different and asymmetric inner vortices are induced, with their cores unevenly displaced in the axial direction, which decrease in shape, become closer to the posts and increase in intensity when $M$ increases. For $M>M_{cr}$, a second external vortex appears in the first quadrant, similar to what happened for $\ell\rightarrow\infty$, but in the case of $\ell=2$, the vortices are deformed and axially enclosed within the semi-distance length. Moreover, in all cases the vortices are also surrounded by a region of streamwise velocity. 

\section{Conclusion}\label{conclusions}
In this work, the problem of oscillatory streams around a cylindrical post, confined between two parallel plates, has been addressed. Eulerian and time-averaged Lagrangian flow fields have been described in detail for $\lambda= \mathcal{O}(1)$ and stroke volumes $\varepsilon < 1$. The problem has been initially tackled analytically and numerically for $\varepsilon \ll 1$, varying the Womersly number, $M$, and the post aspect ratio, $\lambda$.  The numerical results indicate that, when the oscillating flow is harmonic, for a given value of $\lambda$, at low values of $M$ the flow field is sub-critical, showing a recirculating vortex anchored to the post symmetric with respect to the $x$ and $y$ axes. The size of the vortex decreases as $\lambda$ decreases due to the confining effect of the upper and lower walls, at the same time that it narrows and approaches the surface of the post as $M$ increases. However for values of $M$ larger than a critical one, $M_{cr}(\lambda)$, that depends on $\lambda$, the flow becomes super-critical, with the formation of a second, outer vortex also symmetric, whose size decreases as $\lambda$ decreases. The evolution of the critical Womersley number with the post aspect ratio is shown Fig.~\ref{fig:Mcr}, with examples of the flow topology displayed for different values of $\lambda$ and $M$. It can be clearly observed that transition between sub-critical and super-critical regimes is delayed when $\lambda$ decreases, \ie, when the distance between the upper and lower walls is decreased, and the magnitude of the streaming motion is reduced. Regarding the Lagrangian mean motion, given by the sum of the steady-streaming and the Stokes drift, it can be concluded that the Stokes drift prevails over the steady-streaming for sufficiently small values of the Womersley number, whereas in the opposite limit the Stokes-drift motion fades away. Nevertheless, for the range of interest of $M$ in the present work, it has been observed that the steady-streaming  dominats over the Stokes-drift.\\

The numerical results have been complemented by an experimental study, which compares the effect of $M$ in the Lagrangian steady flow patterns at $z = 0$ and for $\lambda = 2$, showing a good agreement. This experimental work has been extended to situations beyond the theory, where $\varepsilon$ is not longer much smaller than one, and cases close to reality, comparing the flow patterns obtained with harmonics and anharmonic waveforms. Furthermore, the mean Lagrangian flow generated by the presence of an array of posts immersed in an oscillatory stream has also been experimentally characterized, emulating the presence of nerve roots in the spinal subarachnoid space. Results obtained for a harmonic flow show time-averaged flow patterns that are identical in all quadrants, in each of which a recirculation cell forms near the post, for $M<M_{cr}$, whose size reduces as $M$ increases. When $M>M_{cr}$, the flow topology changes, and a second, external vortex is formed. The time-averaged flow induced by an anharmonic waveform is found to lose the symmetry respect to the transversal axis, and two vortices of different size are formed at each side of the post in the sub-critical regime ($M< M_{cr}$). However, in the super-critical regime ($M> M_{cr}$) only an outer cell generates in the quadrant downstream, in the systolic direction. Regardless the waveform, the flow topology transition has been found to also depend on the stroke length $\varepsilon$, taking place at values of $M_{cr}$ that decrease as $\varepsilon$ increases. Finally, considering a linear array of posts, the induced flow patterns are similar to those observed with a single post configuration when the stream is harmonic. However, significantly different flow topologies are found for anharmonic waveforms. These results aim to provide a better understanding of the characterization of the influence of microanatomy on CSF flow and lay the foundation for future research, where not only the fluid flow but also the solute transport should be taken into account.
\section*{Acknowledgments}
This work was supported by the coordinated project, PID2020-115961RB-C31, PID2020-115961RB-C32 and PID2020-115961RA-C33, financed by  MICIU/AEI/ 
10.13039/501100011033. F. Moral-Pulido wants to thank the Spanish Ministry of Universities for the financial support provided by the Fellowship FPU18/05694. Fruitful discussions with Prof. Antonio L. Sanchez are greatly acknowledged.
\appendix 
\section{Revisting two-dimensional streaming flow over a circular cylinder}\label{appendix}

Using the same scales as the three dimensional problem for the time $t$, cartesian coordinates $(x,y)$, velocity $\textbf{v}=(u,v)$, and spatial pressure difference $p$ reduces the problem to that of integrating
\begin{equation}
    \nabla\cdot\textbf{v}=0,
\end{equation}
\begin{equation}
    \frac{\partial\textbf{v}}{\partial t}+\varepsilon\textbf{v}\cdot\nabla\textbf{v}=-\nabla p+ \frac{1}{M^2}\nabla^2\textbf{v},
\end{equation}
for $x^2+y^2>1$, subject to the nonslip boundary conditions
\begin{equation}\label{ns_bc}
    \textbf{v}=0 \,\,\
     \textup{at } \,\,\ x^2+y^2=1 \,\,\
\end{equation}

and the far-field condition
\begin{equation}\label{ff_bc}
    \textbf{v}=\left(\textup{cos}t,0\right) \,\,\ \textup{as} \,\,\ x^2+y^2\rightarrow\infty.
\end{equation}
\subsection{Eulerian velocity for $\varepsilon\ll 1$}
Following standard practice, we describe the flow by introducing expansions for the
different flow variables in powers of $\varepsilon$, i.e.
\begin{equation}
    \textbf{v}=\textbf{v}_0+\varepsilon\textbf{v}_1+...
\end{equation}
and $p=p_0+\varepsilon p_1+...$.
The leading-order solution satisfies the linear equation
\begin{equation}\label{lo_continuity}
    \nabla\cdot\textbf{v}_0=0,
\end{equation}
\begin{equation}\label{lo_momentum}
    \frac{\partial\textbf{v}_0}{\partial t}=-\nabla p_0+ \frac{1}{M^2}\nabla^2\textbf{v}_0,
\end{equation}
subject to boundary conditions stated in (\ref{ns_bc}) and (\ref{ff_bc}). The problem can be alternatively written using $\textbf{v}_0=\textup{Re}(e^{it}\textbf{V}_0)$ with $\textbf{V}_0(x,y)=(U_0,V_0)$ representing a complex velocity. The integration can make use of the stream function $\Psi_0(x,y)$ defined such that
\begin{equation}
    U_0=\frac{\partial\Psi_0}{\partial y} \,\,\ \textup{and} \,\,\ V_0=-\frac{\partial\Psi_0}{\partial x}
\end{equation}
and
\begin{equation}\label{lo_psi}
    \Omega_0=-\nabla^2\Psi_0,
\end{equation}
where the complex function $\Omega_0(x,y)$, related to the vorticity by
\begin{equation}
    \omega_0=\frac{\partial v_0}{\partial x}-\frac{\partial u_0}{\partial y}=\textup{Re}(e^{it}\Omega_0),
\end{equation}
satisfies
\begin{equation}\label{lo_vorticity}
    iM^2\Omega_0=\nabla^2\Omega_0,
\end{equation}
as follows from (\ref{lo_momentum}). Equations (\ref{lo_psi}) and (\ref{lo_vorticity}) must be integrated subject to
\begin{equation}\label{noslip_psi0}
    \Psi_0=(x,y)\cdot\nabla\Psi_0=0 \,\,\ \textup{at} \,\,\ \textup{$x^2+y^2=1$}, 
\end{equation}
\begin{equation}\label{farfield_psi0}
    \frac{\partial\Psi_0}{\partial y}-1=\frac{\partial\Psi_0}{\partial x}=0 \,\,\ \textup{as} \,\,\ \textup{$x^2+y^2\rightarrow\infty$}.
\end{equation}
For a general value of $M$, the resulting velocity $\textbf{V}_0(x,y)=(U_0,V_0)$ has real and imaginary
parts. Note, however, that in the limit of steady creeping flow $M\ll1$ the solution is
real everywhere, while in the inviscid limit $M\gg1$ the solution contains an imaginary
part only in the thin Stokes layer of thickness $1/M$ hat develops on the cylinder surface, outside of which $\Omega_0=0$ and $\textup{Im}(\textbf{V}_0=0)$.\\

Formulating the problem in cylindrical polar coordinates instead of Cartesians and separating variables, i.e. $\Omega_0(r,\theta)=\Omega_R(r)\Omega_{\Theta}(\theta)$, we find the following solution of the equation (\ref{lo_vorticity})
\begin{equation}\label{vorticity_0}
    \Omega_0(r,\theta)=\frac{\pi}{2}\Omega iH_1^{(1)}\left(\alpha M r\right)\sin\theta,
\end{equation}

where $\Omega$ is a constant to be determined by the boundary conditions, $H_1^{(1)}$ is the Hankel function of first kind and order one and $\alpha=\sqrt{i}$. \\

Using the fact that the equation (\ref{lo_psi}) is linear, it can be applied the superposition principle. 
\begin{equation}
    \nabla^2\Psi_0=-\Omega_0:\,\ \left\lbrace\begin{array}{ll}
    \nabla^2\hat{\Psi}_0=0\,\,\ \textup{with b.c}\,\,\ (\ref{noslip_psi0})-(\ref{farfield_psi0}) \\
    \nabla^2\tilde{\Psi}_0=-\Omega_0\,\,\     \textup{with homogeneous b.c.}
    \end{array}
\right.
\end{equation}
The solution of $\nabla^2\hat{\Psi}_0=0$ with the boundary conditions (\ref{noslip_psi0}) and (\ref{farfield_psi0}) can be expressed as
\begin{equation}
    \hat{\Psi}_0=A\left(r+\frac{B}{r}\right)\sin\theta,
\end{equation}
that has the same form as potential solution for the flow. The value of the constants of integration $A$ and $B$ will be determined below.\\ 

The equation $\nabla^2\tilde{\Psi}_0=-\Omega_0$ is solved using the Frobenius method, yielding:
\begin{equation}
    \tilde{\Psi}_0=\frac{1}{\alpha M}\frac{H_1^{(1)}\left(\alpha M r\right)}{H_0^{(1)}\left(\alpha M\right)}\sin\theta
\end{equation}

The leading order stream function will be:
\begin{equation}
    \Psi_0=\hat{\Psi}_0+\tilde{\Psi}_0.
\end{equation}
In this case, the stream function will be:

\begin{equation}
    \Psi_0=\left[\frac{1}{\alpha M}\frac{H_1^{(1)}\left(\alpha M r\right)}{H_0^{(1)}\left(\alpha M\right)}-\frac{1}{2} \left(r+\frac{1}{r}\frac{H_2^{(1)}\left(\alpha M\right)}{H_0^{(1)}\left(\alpha M\right)}\right)\right]\sin\theta
\end{equation}
To calculate the constant $\Omega$ it must be imposed the no slip boundary condition for the azimuthal velocity ($v_{\theta0}=-\frac{\partial\Psi_0}{\partial r}=0$ at $r=1$). The value of the constant $\Omega$ in the equation (\ref{vorticity_0}) is:
\begin{equation}
    \Omega=\frac{H_2^{(1)}\left(\alpha M\right)}{H_0^{(1)}\left(\alpha M\right)}.
\end{equation}

As follows from collecting terms of order $\varepsilon$, the problem at the following order becomes
\begin{equation}\label{v1_continuity}
    \nabla\cdot\textbf{v}_1=0,
\end{equation}
\begin{equation}\label{v1_momentum_app}
    \frac{\partial\textbf{v}_1}{\partial t}+\textbf{v}_0\cdot\nabla\textbf{v}_0=-\nabla p_1+ \frac{1}{M^2}\nabla^2\textbf{v}_1,
\end{equation}
with boundary conditions
\begin{equation}\label{v1_bc}
    \textbf{v}_1=0 \left\lbrace \begin{array}{ll}
     \textup{at } \,\,\ x^2+y^2=1\\
     \textup{as } \,\,\ x^2+y^2\to\infty
    \end{array}
    \right.
\end{equation}
The first-order corrections $\textbf{v}_1$ and $p_1$ are $2\pi$-peridic functions of time. As a result of the nonlinear interactions associated with the convective terms, their time-averaged values $\langle\textbf{v}_1\rangle$ and $\langle p_1\rangle$ are nonzero, with $\langle\cdot\rangle=\int_0^{2\pi}\cdot\textup{d}t$ denoting the time-
average operator. Taking the time average of (\ref{v1_continuity})-(\ref{v1_bc}) and using the identity $\langle\textbf{v}_0\cdot\nabla\textbf{v}_0\rangle=\frac{1}{2}\textup{Re}(\textbf{V}_0\cdot\nabla\textbf{V}_0^{*})$, where the asterisk $*$ denotes complex conjugates, provides the
steady-streaming problem
\begin{equation}\label{v1aver_continuity}
    \nabla\cdot\langle\textbf{v}_1\rangle=0,
\end{equation}
\begin{equation}\label{v1aver_momentum}
    \frac{1}{2}\textup{Re}(\textbf{V}_0\cdot\nabla\textbf{V}_0^{*})=-\nabla \langle p_1\rangle+ \frac{1}{M^2}\nabla^2\langle\textbf{v}_1\rangle,
\end{equation}
with boundary conditions
\begin{equation}\label{v1aver_bc}
    \langle\textbf{v}_1\rangle=0 \left\lbrace \begin{array}{ll}
     \textup{at } \,\,\ x^2+y^2=1\\
     \textup{as } \,\,\ x^2+y^2\to\infty
    \end{array}
    \right.
\end{equation}
Introducing the time-averaged stream function $\langle\psi_1\rangle$, with
\begin{equation}
    \langle u_1\rangle=\frac{\partial\langle\psi_1\rangle}{\partial y}\,\,\ \textup{and} \,\,\ \langle v_1\rangle=-\frac{\partial\langle\psi_1\rangle}{\partial x}
\end{equation}
and associated vorticity $\langle\omega_1\rangle=-\nabla^2\langle\psi_1\rangle$, reduces (\ref{v1aver_momentum}) to
\begin{equation}\label{psi1_momentum}
    -\frac{M^2}{2}\textup{Re}\left(\frac{\partial\Psi_0}{\partial y}\frac{\partial\Omega_0^{*}}{\partial x}-\frac{\partial\Psi_0}{\partial x}\frac{\partial\Omega_0^{*}}{\partial y}\right)=\nabla^2\nabla^2\langle\psi_1\rangle
\end{equation}
involving the complex functions $\Psi_0$ and $\Omega_0^{*}=-\nabla^2\Psi_0^{*}$. The resulting recirculating cells can be obtained by integrating (\ref{psi1_momentum}) subject to the
boundary conditions:
\begin{equation}\label{app.bc.ss1}
   \langle\psi_1\rangle=(x,y)\cdot\nabla\langle\psi_1\rangle=0 \,\,\ \textup{at} \,\,\ \textup{$x^2+y^2=1$}, 
\end{equation}
\begin{equation}\label{app.bc.ss2}
     \langle\psi_1\rangle=0 \,\,\ \textup{as} \,\,\ \textup{$x^2+y^2\rightarrow\infty$}.
\end{equation}
In cylindrical polar coordinates:
\begin{equation}\label{psi1_momentum_p}
    -\frac{M^2}{2}\textup{Re}\left[\frac{1}{r}\left(\frac{\partial\Psi_0}{\partial \theta}\frac{\partial\Omega_0^{*}}{\partial r}-\frac{\partial\Psi_0}{\partial r}\frac{\partial\Omega_0^{*}}{\partial \theta}\right)\right]=\nabla^2\nabla^2\langle\psi_1\rangle
\end{equation}
The equation can be solved using the method of separation of variables, supposing $\langle\psi_1\rangle=\psi_{1R}(r)\sin2\theta$, and the method of variations of parameters to do:

\begin{equation}
\begin{array}{cc}
      \psi_{1R}(r) &= r^4\left(\frac{1}{48}\int_1^{r}\frac{1}{r'}\Phi(r')\textup{d}r' + C_1\right) -r^2\left(\frac{1}{16}\int_1^{r}r'\Phi(r')\textup{d}r' + C_2\right)+ \\
     & \left(\frac{1}{16}\int_1^{r}r'^{3}\Phi(r')\textup{d}r' + C_3\right)+\frac{1}{r^2}\left(-\frac{1}{48}\int_1^{r}r'^{5}\Phi(r')\textup{d}r' + C_4\right),
\end{array}
\end{equation}

where
\begin{equation}
\Phi(r)=iM^2\textup{Im}\left[\frac{H_2^{(1)}(\alpha M r)}{H_0^{(1)}(\alpha M)}+\frac{2}{\left\|H_0^{(1)}(\alpha M)\right\|^2}\left(\frac{1}{r^2}H_2^{(1)}(\alpha M)H_0^{(1)}(\alpha r)+H_2^{(1)}(\alpha M r)H_0^{(1)}(\alpha M )\right)\right],
\end{equation}    
and $$\left\|H_0^{(1)}(\alpha M)\right\|^2=\textup{Re}(H_0^{(1)}(\alpha M))^2+\textup{Im}(H_0^{(1)}(\alpha M))^2.$$
The constants
\begin{equation}
\begin{array}{lll}
     C_1 = -\frac{1}{48}\int_1^{\infty}\frac{1}{r}\Phi(r)\textup{d}r, \,\ C_2 = \frac{1}{16}\int_1^{\infty}r\Phi(r)\textup{d}r,\\
     C_3 = \frac{1}{16}\int_1^{\infty}\frac{1}{r}\Phi(r)\textup{d}r -\frac{1}{8}\int_1^{\infty}r\Phi(r)\textup{d}r,\\
     C_4 = -\frac{1}{24}\int_1^{\infty}\frac{1}{r}\Phi(r)\textup{d}r + \frac{1}{16}\int_1^{\infty}r\Phi(r)\textup{d}r, 
    \end{array}
\end{equation}
are determined by imposing the boundary conditions \ref{app.bc.ss1},\ref{app.bc.ss2}. The Lagrangian mean motion can be defined as:
\begin{equation}   \textbf{v}_L=\langle\textbf{v}_1\rangle+\frac{1}{2}\textup{Im}(\textbf{V}_0\cdot\nabla\textbf{V}_0^{*})
\end{equation}
where $\frac{1}{2}\textup{Im}(\textbf{V}_0\cdot\nabla\textbf{V}_0^{*})$ is the Stokes drift or correction. In the present case, the Streamfunction of the Stokes drift can be calculated as:
\begin{equation}\label{v_sd1}
   \psi_{SD}=\frac{1}{2}\textup{Im}\left[\frac{H_2^{(1)}(\alpha Mr)}{H_0^{(1)}(\alpha M)}+\frac{1}{r^2}\frac{H_2^{(1)}(\alpha M)}{H_0^{(1)}(\alpha M)}\left(\frac{H_0^{(1)*}(\alpha Mr)}{H_0^{(1)*}(\alpha M)}-1\right)+\frac{H_0^{(1)}(\alpha Mr)}{H_0^{(1)}(\alpha M)}\frac{H_2^{(1)*}(\alpha Mr)}{H_0^{(1)*}(\alpha M)}\right]\sin2\theta.
\end{equation}\\

\section*{Declaration of interest.} The authors report no conflict of interest.

\bibliographystyle{unsrtnat}

%\bibliography{post_bibliography}

\begin{thebibliography}{39}
\providecommand{\natexlab}[1]{#1}
\providecommand{\url}[1]{\texttt{#1}}
\expandafter\ifx\csname urlstyle\endcsname\relax
  \providecommand{\doi}[1]{doi: #1}\else
  \providecommand{\doi}{doi: \begingroup \urlstyle{rm}\Url}\fi

\bibitem[Linninger et al.(2016)]{Linninger2016}
A.A. Linninger, K. Tangen, C.Y. Hsu, and D. Frim. 
\newblock Cerebrospinal fluid mechanics and its coupling to cerebrovascular dynamic
\newblock \emph{Annu. Rev. Fluid Mech.}, 48\penalty0 (1): \penalty0 219–-257, 2016.

\bibitem[Grotberg and Jensen(2004)]{Grotberg2004}
J.B. Grotberg and O.E. Jensen.
\newblock Biofluid mechanics in flexible tubes.
\newblock \emph{Annu. Rev. Fluid Mech.}, 36\penalty0 (1):\penalty0 121--147, 2004.

\bibitem[Kelley et~al.(2023)]{Kelley2023}
D.H. Kelley and J.H. Thomas.
\newblock Cerebrospinal fluid flow.
\newblock \emph{Annu. Rev. Fluid Mech.}, 55:\penalty0 2237–26, 2023.

\bibitem[Dauleac et~al.(2019)]{Dauleac2019}
C. Dauleac, T. Jacquesson, and P. Mertens.
\newblock Anatomy of the human spinal cord arachnoid cisterns: applications for spinal cord surgery. 
\newblock \emph{J. Neurosurg.}, 31 \penalty0 (5): \penalty0 756–-763, 2019.

\bibitem[Stockman (2006)]{stockman2006effect}
H.W. Stockman. 
\newblock Effect of anatomical fine structure on the flow of cerebrospinal fluid in the spinal subarachnoid space.
\newblock \emph{J. Biomech. Eng.}, 128\penalty0 (1):\penalty0 106–-114, 2006.

\bibitem[Stockman (2007)]{stockman2007effect}
H.W. Stockman. 
\newblock Effect of anatomical fine structure on the dispersion of solutes in the spinal subarachnoid space.
\newblock \emph{J. Biomech. Eng.}, 128\penalty0 (5):\penalty0 666-–675, 2007.

\bibitem[Tangen et al.(2015)]{tangen2015}
K.M. Tangen, Y. Hsu, D.C. Zhu, and A.A. Linninger. 
\newblock NS wide simulation of flow resistance and drug transport due to spinal microanatomy.
\newblock \emph{J. Biomech.}, 48\penalty0 (10):\penalty0 2144–-2154, 2015.

\bibitem[ Pahlavia et~al.(2014)]{pahlavian2014impact}
S.H. Pahlavian, T. Yiallourou, R.S. Tubbs, A.C. Bunck, F. Loth, M. Goodin, M. Raisee, and B.A. Martin.
\newblock The impact of
spinal cord nerve roots and denticulate ligaments on cerebrospinal fluid dynamics in the cervical spine.
\newblock \emph{PLoS One}, 9\penalty0 (4):\penalty0 e91888, 2014.

\bibitem[Reina et~al.(2020)]{Reina2020}
M.A. Reina, A. Boezaart, C. De Andr\'es-Serrano, R. Rubio-Haro, and J. De Andr\'es
\newblock Drug Delivery Systems. Methods in Molecular Biology,
\newblock volume 2059.\emph{Humana},2020.

\bibitem[Ayansiji et~al.(2023)]{Ayansiji2023determination}
A.O. Ayansiji, D.S. Gehrke, B. Baralle, Ariel Nozain, M.R. Singh, and A.A. Linninger. 
\newblock Determination of spinal tracer dispersion after intrathecal injection in a deformable CNS model.
\newblock \emph{Frontiers in Physiology}, 14\penalty0, 2023.

\bibitem[Lawrence et~al.(2019)]{paper2}
J. J. Lawrence, W. Coenen, A. L. S\'anchez, G. Pawlak, C. Mart\'inez-Baz\'an, V. Haughton, and J.C. Lasheras. 
\newblock  On the dispersion of a drug delivered intrathecally in the spinal canal.
\newblock \emph{J. Fluid Mech}, 869:\penalty0 679–-720, 2019.

\bibitem[Riley (2001)]{riley2001steady}
N. Riley.
\newblock Steady streaming.
\newblock \emph{Ann. Rev. Fluid Mech.}, 33\penalty0 (1):\penalty0 43-–65, 2001.

\bibitem[S\'anchez et~al.(2018)S\'anchez, Mart\'inez-Baz\'an, Guti\'errez-Montes, Criado-Hidalgo, Pawlak, Bradley, Haughton, and Lasheras]{paper1}
A.~L. S\'anchez, C.~Mart\'inez-Baz\'an, C.~Guti\'errez-Montes, E.~Criado-Hidalgo, G.~Pawlak, W.~Bradley, V.~Haughton, and J.~C. Lasheras.
\newblock On the bulk motion of the cerebrospinal fluid in the spinal canal.
\newblock \emph{J. Fluid Mech.}, 841:\penalty0 203--227, 2018.


\bibitem[Stokes (1847)]{Stokes1847}
G.G. Stokes.
\newblock On the theory of oscillating waves.
\newblock \emph{Trans. Camb. Phil. Soc}, 8:\penalty0 441–-455, 1847.

\bibitem[Rayleigh (1884)]{Rayleigh1883}
Lord Rayleigh.
\newblock On the circulation of air observed in kundt’s tubes, and on some allied acoustical problems
\newblock \emph{ Phil. Trans. Royal Soc. London}, 175:\penalty0 1--21, 1884.

\bibitem[Holtsmark et~al.(1954)]{holtsmark1954boundary}
J. Holtsmark, I. Johnsen, T. Sikkeland, and S. Skavlem.
\newblock Boundary layer flow near a cylindrical obstacle in an oscillating, incompressible fluid 
\newblock \emph{J. Acoust. Soc. Am.}, 26\penalty0 (1):\penalty0 26--39, 1954.


\bibitem[Chong et al.(2013)]{chong2013inertial}
K. Chong, S.D Kelly, S. Smith, and J.D. Eldredge.
\newblock Inertial particle trapping in viscous streaming.
\newblock \emph{Phys. Fluids}, 25\penalty0 (3):\penalty0 033602, 2013.

\bibitem[Raney et al.(1954)]{raney1954acoustical}
W.P. Raney, J.C. Corelli, and P.J. Westervelt.
\newblock Acoustical streaming in the vicinity of a cylinder.
\newblock \emph{J. Acoust. Soc. Am.}, 26\penalty0 (6):\penalty0 1006–-1014, 1954.

\bibitem[Skavlem and Tj\"otta (1955)]{skavlem1955coaxial}
S. Skavlem and S. Tj\"otta.
\newblock Steady rotational flow of an incompressible, viscous fluid enclosed between two coaxial cylinders.
\newblock \emph{J. Acoust. Soc. Am.}, 27\penalty0 (1):\penalty0 26–-33, 1955.

\bibitem[Tatsuno (1973)]{tatsuno1973}
M. Tatsuno.
\newblock Circulatory streaming around an oscillating circular cylinder at low Reynolds numbers.
\newblock \emph{J.Phys. Soc. Japan}, 35\penalty0 (2):\penalty0 915-–920, 1973.

\bibitem[Tatsuno and Bearman (1990)]{tatsuno_bearman_1990}
M. Tatsuno and P.W. Bearman.
\newblock A visual study of the flow around an oscillating circular cylinder at low keulegan–carpenter
numbers and low stokes numbers.
\newblock \emph{J. Fluid Mech.}, 211:\penalty0 157-–182, 1990.

\bibitem[Justesen (1991)]{justesen_1991}
P. Justesen.
\newblock  A numerical study of oscillating flow around a circular cylinder.
\newblock \emph{J. Fluid Mech.}, 222:\penalty0 157-–196, 1991.

\bibitem[Lutz et al.(2005)]{lutz2005microscopic}
B. R. Lutz, J. Chen, and D. T. Schwartz.
\newblock Microscopic steady streaming eddies created around short cylinders in a channel:619
Flow visualization and stokes layer scaling.
\newblock \emph{Phys. Fluids}, 17(2):\penalty0 0023601, 2005.

\bibitem[Lutz et al.(2006)]{lutz2006}
B. R. Lutz, J. Chen, and D. T. Schwartz.
\newblock Hydrodynamic tweezers: 1. noncontact trapping of single cells using steady streaming
microeddies. 
\newblock \emph{Anal. Chem.}, 78(15):\penalty0 5429–-5435, 2006.

\bibitem[Tien et al.(2013)]{tien2013}
W. Tien, D. Dabiri, V.H. Lieu, and D.T. Schwartz.
\newblock Volumetric velocity measurements of an acoustic streaming microeddy array using color-coded three-dimensional micro particle tracking velocimetry.
\newblock \emph{10TH Int. Symp PIV – PIV13, Delft},  2013, july\penalty0 2--4.

\bibitem[Marin et al.(2015)]{marin2015three}
A. Marin, M. Rossi, B. Rallabandi, C. Wang, Hilgenfeldt S., and C.J. K\"ahler. 
\newblock Three-dimensional phenomena in microbubble
acoustic streaming.
\newblock \emph{Phys. Rev. Applied},  3(041001):\penalty0 1–-5, 2015.

\bibitem[Rallabandi et al.(2015)]{rallabandi2015three}
B. Rallabandi, A. Marin, M. Rossi, C.J. K\"ahler, and S. Hilgenfeldt. 
\newblock Three-dimensional streaming flow in confined geometries.
\newblock \emph{J. Fluid Mech.},  777:\penalty0 408–-429, 2015.


\bibitem[Chan et al.(2022)]{chan_2022}
F. K. Chan, Y. Bhosale, T. Parthasarathy, and M. Gazzola.
\newblock Three-dimensional geometry and topology effects in viscous streaming.
\newblock \emph{J. Fluid Mech.}, 933:\penalty0 A53, 2022.

\bibitem[Coenen (2016)]{coenen2016}
W. Coenen.
\newblock Steady streaming around a cylinder pair.
\newblock \emph{Proc. R. Soc. A: Math. Phys. Eng. Sci.},  472(2195), 2016.

\bibitem[Alaminos-Quesada et al. (2023a)]{Alaminos-Quesada2023}
J. Alaminos-Quesada, J. J. Lawrence, W. Coenen, and A. L. S\'anchez.
\newblock  Oscillating viscous flow past a streamwise linear array of circular cylinders.
\newblock \emph{J. Fluid Mech.},  959(A39), 2023.

\bibitem[Liu et al. (2015)]{Liu2015}
Y. Liu, X. Zhou, J. Ma, Y. Ge, and X. T. Cao.
\newblock  The diameters and number of nerve fibers in spinal nerve roots.
\newblock \emph{J. Spinal Cord Med.},  38(4): 532--537, 2015.

\bibitem[Coenen et al. (2019)]{coenen2019subject}
W. Coenen, C. Gutierrez-Montes, S. Sincomb, E. Criado-Hidalgo, K. Wei, K. King, V. Haughton, C. Mart\'inez-Baz\'an, A.L. S\'anchez, and J.C. Lasheras.
\newblock Subject-specific studies of csf bulk flow patterns in the spinal canal: implications for the dispersion of solute particles in intrathecal drug delivery.
\newblock \emph{AJNR Am. J. Neuroradiol.},  40(7): 1242–-1249, 2019.

\bibitem[Sass et al. (2017)]{sass20173d}
L.R. Sass, M. Khani, G. Natividad, R.S. Tubbs, O. Baledent, and B.A. Martin.
\newblock A 3D subject-specific model of the spinal subarachnoid space with anatomically realistic ventral and dorsal spinal cord nerve rootlets.
\newblock \emph{ Fluids and Barriers of the CNS,},  14(1): 36, 2017.

\bibitem[Jeong and Hussain (1995)]{jeong_hussain_1995}
J. Jeong and F. Hussain.
\newblock On the identification of a vortex.
\newblock \emph{J. Fluid Mech.},  285: 69--94, 1995.

\bibitem[Katsanoulis et al. (2023)]{katsanoulis_2023}
S. Katsanoulis, F. Kogelbauer, R. Kaundinya, J. Ault, and G. Haller.
\newblock On the identification of a vortex.
\newblock \emph{J. Fluid Mech.},  954:A28, 2023.

\bibitem[Thielicke and Stamhuis (2014)]{Stamhuis2014}
W. Thielicke and E. Stamhuis.
\newblock Pivlab-towards user-friendly, affordable and accurate digital particle image velocimetry in
matlab.
\newblock \emph{J. Open Res. Softw.},  2(1), 2014.

\bibitem[Alaminos-Quesada (2021)]{alaminos2021}
J. Alaminos-Quesada.
\newblock Limit of the two-dimensional linear potential theories on the propulsion of a flapping airfoil in forward flight in terms of the reynolds and strouhal number. 
\newblock \emph{Phys. Rev. Fluids},  6:123101, 2021.

\bibitem[Sincomb et al. (2022)]{Sincomb2022}
S. Sincomb, W. Coenen, C. Guti\'errez-Montes, C. Mart\'inez-Baz\'an, V. Haughton, and A. L. S\'anchez.
\newblock A one-dimensional model for the pulsating flow of cerebrospinal fluid in the spinal canal.
\newblock \emph{J. Fluid Mech.},  939:A26, 2022.

\bibitem[Davidson and Riley (1972)]{Davidson1972}
B.J. Davidson and N. Riley.
\newblock Jets induced by oscillatory motion.
\newblock \emph{J. Fluid Mech.},  53(2):287--303, 1972.

\bibitem[Alaminos-Quesada et al. (2023b)]{alaminosquesadaAMMOD2023}
J. Alaminos-Quesada, C. Guti\'errez-Montes, W. Coenen and A. L. S\'anchez.
\newblock Stationary flow driven by non-sinusoidal time-periodic pressure gradients in wavy-walled channels. 
\newblock \emph{Appl. Math. Model.},   122:693–705, 2023.

\bibitem[Tatsuno (1981)]{tatsuno1981secondary}
M. Tatsuno.
\newblock Secondary flow induced by a circular cylinder performing unharmonic oscillations.
\newblock \emph{ J. Phys. Soc. Japan},   50(1):330–337, 1981.

\end{thebibliography}

\end{document}